\documentclass[utf8]{FrontiersinHarvard} 

\usepackage{listings, xcolor}

\definecolor{verylightgray}{rgb}{.97,.97,.97}

\lstdefinelanguage{Solidity}{
	keywords=[1]{anonymous, assembly, assert, balance, break, call, callcode, case, catch, class, constant, continue, constructor, contract, debugger, default, delegatecall, delete, do, else, emit, event, experimental, export, external, false, finally, for, function, gas, if, implements, import, in, indexed, instanceof, interface, internal, is, length, library, log0, log1, log2, log3, log4, memory, modifier, new, payable, pragma, private, protected, public, pure, push, require, return, returns, revert, selfdestruct, send, solidity, storage, struct, suicide, super, switch, then, this, throw, transfer, true, try, typeof, using, value, view, while, with, addmod, ecrecover, keccak256, mulmod, ripemd160, sha256, sha3}, 
	keywordstyle=[1]\color{black}\bfseries,
	keywords=[2]{address, bool, byte, bytes, bytes1, bytes2, bytes3, bytes4, bytes5, bytes6, bytes7, bytes8, bytes9, bytes10, bytes11, bytes12, bytes13, bytes14, bytes15, bytes16, bytes17, bytes18, bytes19, bytes20, bytes21, bytes22, bytes23, bytes24, bytes25, bytes26, bytes27, bytes28, bytes29, bytes30, bytes31, bytes32, enum, int, int8, int16, int24, int32, int40, int48, int56, int64, int72, int80, int88, int96, int104, int112, int120, int128, int136, int144, int152, int160, int168, int176, int184, int192, int200, int208, int216, int224, int232, int240, int248, int256, mapping, string, uint, uint8, uint16, uint24, uint32, uint40, uint48, uint56, uint64, uint72, uint80, uint88, uint96, uint104, uint112, uint120, uint128, uint136, uint144, uint152, uint160, uint168, uint176, uint184, uint192, uint200, uint208, uint216, uint224, uint232, uint240, uint248, uint256, var, void, ether, finney, szabo, wei, days, hours, minutes, seconds, weeks, years},	
	keywordstyle=[2]\color{black}\bfseries,
	keywords=[3]{block, blockhash, coinbase, difficulty, gaslimit, number, timestamp, msg, data, gas, sender, sig, value, now, tx, gasprice, origin},	
	keywordstyle=[3]\color{black}\bfseries,
	identifierstyle=\color{black},
	sensitive=true,
	comment=[l]{//},
	morecomment=[s]{/*}{*/},
	commentstyle=\color{gray}\ttfamily,
	stringstyle=\color{black}\ttfamily,
	morestring=[b]',
	morestring=[b]"
}

\usepackage{url,lineno,microtype,subcaption}
\usepackage[dvipsnames]{xcolor}
\usepackage[pdftex, colorlinks=true, hyperfootnotes=true, hyperindex=true,
plainpages=false, pagebackref=false, pdfpagelabels=true, pdfstartview=FitH,
linkcolor=purple, citecolor=violet, urlcolor=blue,
bookmarks, bookmarksopen, bookmarksdepth=3]{hyperref}
\usepackage[onehalfspacing]{setspace}
\usepackage{listings}
\usepackage{todonotes}
\usepackage{tabu}
\usepackage{slashbox}
\usepackage{comment}
\usepackage{paralist}
\usepackage[capitalise,nameinlink]{cleveref}
\usepackage[scientific-notation=false,group-separator={,}]{siunitx}
\usepackage[T1]{fontenc}
\usepackage[justification=centering]{caption}
\usepackage{titlecaps}

\newcommand{\subsubsubsection}[1]{\noindent\textbf{#1}}

\newcommand{\Compo}[1]{%
  \texttt{
    \titlecap{
      #1}
  }
}

\newenvironment{newt}[1][]{%
	\color{black}%
 
}{%
	\normalcolor%
}
\newcommand{\Newt}[1]{#1}

\def\keyFont{\fontsize{8}{11}\helveticabold }
\def\firstAuthorLast{Basile {et~al.}} 
\def\Authors{Davide Basile\,$^{1}$, Claudio Di Ciccio\,$^{1}$, Valerio Goretti\,$^{1}$ and Sabrina Kirrane\,$^{2}$}

\def\Address{$^{1}$Sapienza University of Rome, Italy\\
$^{2}$Vienna University of Economics and Business, Austria}

\def\corrAuthor{Valerio Goretti}
\def\corrEmail{valerio.goretti@uniroma1.it}

\begin{document}
\onecolumn
\firstpage{1}

\title[Blockchain based Resource Governance for Decentralized Web Environments]{Blockchain based Resource Governance for Decentralized Web Environments} 

\author[\firstAuthorLast ]{\Authors} 
\address{} 
\correspondance{} 

\extraAuth{}

\maketitle

\begin{abstract}
\label{sec:abstract}
\Newt{Decentralization initiatives such as Solid, Digi.me, and ActivityPub aim to give data owners more control over their data and to level the playing field by enabling small companies and individuals to gain access to data, thus stimulating innovation. However, these initiatives typically employ access control mechanisms that cannot verify compliance with usage conditions after access has been granted to others. In this paper, we extend the state of the art by proposing a resource governance conceptual framework, entitled ReGov, that facilitates usage control in decentralized web environments. We subsequently demonstrate how our framework can be instantiated by combining blockchain and trusted execution environments. Through blockchain technologies, we record policies expressing the usage conditions associated with resources and monitor their compliance. Our instantiation employs trusted execution environments to enforce said policies, inside data consumers' devices.} We evaluate the framework instantiation through a detailed analysis of requirements derived from a data market motivating scenario, as well as an assessment of the security, privacy, and affordability aspects of our proposal. 
\tiny
 \keyFont{ \section{Keywords:} Decentralization; Usage Control; Governance; Blockchain;  Trusted Execution Environment} 
\end{abstract}

%
%
\section{Introduction}
\label{sec:introduction}
Since its development, the internet has steadily evolved into a ubiquitous ecosystem that is seen by many as a public utility \citep{quail2010net}. The development of centralized web-based platforms on top of the internet has undoubtedly brought benefits from both an economic and a social perspective. However, the web as we know it today, is dominated by a small number of stakeholders that have a disproportionate influence on the content that the public can produce and consume. The scale of the phenomenon has brought about the need for legal initiatives aimed at safeguarding content producer rights \citep{quintais2020new}. In parallel, technical decentralization initiatives such as Solid\footnote{\url{https://solidproject.org/about}. Accessed: \today.}, Digi.me\footnote{\url{https://digi.me/what-is-digime/}. Accessed: \today.}, and 
ActivityPub\footnote{\url{https://activitypub.rocks/}. Accessed: \today} aim to give data owners more control over their data, while at the same time providing small companies as well as individuals with access to data, which is usually monopolized by centralized platform providers, thus stimulating innovation. To this end, the Solid community are developing tools, best practices, and web standards that facilitate ease of data integration and support the development of decentralized social applications based on Linked Data principles. In turn, Digi.me are developing tools and technologies that enable individuals to download their data from centralized platforms such that they can store it in an encrypted personal data store and leverage a variety of applications that can process this data locally on the data owners device. These client-side applications are developed by innovative app developers who use the Digi.me software development kit to communicate with the encrypted personal data stores directly. Following the same principles, ActivityPub is a decentralized social networking protocol, published by the W3C Social Web Working Group that offers a client-server application programming interface (API) for adding, modifying, and removing material as well as a federated server-server API for sending notifications and subscribing to content. Social networks implementing ActivityPub can be easily integrated with each other in order to form a larger ecosystem, commonly referred to as the Fediverse\footnote{\url{https://fediverse.party/en/fediverse/}. Accessed: \today.}. Some of the most popular Fediverse initiatives include Mastodon\footnote{\url{https://docs.joinmastodon.org}. Accessed: \today.}, PeerTube\footnote{\url{https://peertube.uno}. Accessed: \today.}, and PixelFed\footnote{\url{https://pixelfed.uno/site/about}. Accessed: \today.}.

In order to better cater for use case scenarios that involve data sharing across various distributed data stores underpinning decentralized applications, there is a need for tools and technologies that are not only capable of working with distributed data but are also able to manage data resources that come with a variety of usage terms and conditions specified by data producers. However, the vast majority of decentralized web initiatives, which aim to provide users with a greater degree of control over personal resources,  manage data access via simple access control mechanisms \citep{ouaddah2016fairaccess,toninelli2006semantic,tran2005trust} that are not able to verify that usage conditions are adhered to after access has been granted \citep{akaichi2022usage}. \Newt{For example, access control rules can determine if users can retrieve data or not. However, they cannot express conditions on the type of application that can process them, the geographical area in which they can be treated, when the access grant would expire, or the number of times they can be processed.} 

\Newt{When it comes to the realization of usage control in decentralized web environments, Trusted Execution Environments (TEEs) and Distributed Ledger Technologies (DLTs) could serve as fundamental enablers. Trusted execution environments offer data and code integrity to enforce the conditions established by decentralized data providers, directly in consumers' devices. DLTs can store shared policies in a distributed ecosystem in which data usage is governed by smart contracts, while recording an immutable log of usage operations.} 

To this end, in this paper we propose a resource governance (ReGov) conceptual framework and an instantiation thereof. ReGov combines blockchain applications and trusted execution environments to facilitate usage control in decentralized web environments.
The work is guided by a typical decentralized web scenario, according to which data are not stored in centralized servers but rather in decentralized data stores controlled by users. Throughout the paper, we refer to the component for managing the data stored locally on every user's device as a \emph{data node} (or \emph{node} for simplicity). 

In terms of contributions, we extend the state of the art by: (i) proposing a generic resource governance conceptual framework; (ii) demonstrating how blockchain technologies and trusted execution environments can together be used to manage resource usage; and (iii) assessing the effectiveness of the proposed framework via concrete quantitative and qualitative evaluation metrics derived from our data market motivating use case scenario.

The remainder of the paper is structured as follows: \cref{sec:background} presents the necessary background information regarding data access and usage control, trusted execution environments, decentralized applications, and blockchain oracles. In the same section we also provide an overview of related work. We introduce the motivating scenario used to guide our work in \cref{sec:motivation} and our ReGov conceptual framework in \cref{sec:framework}. Following on from this, we described our DLT and TEE-based instantiation in \cref{sec:instantiation} and the results of our quantitative and qualitative in \cref{sec:evaluation}. Finally, we conclude and outline directions for future work in \cref{sec:conclusion}.

%
%
\section{Background and Related Work}
\label{sec:background}

\begin{newt}
This section sets the context for the work being presented, highlighting the significance and relevance of the study. It also gives credit to previous work in the field and identifies gaps in the current understanding that the study aims to fill.
\end{newt}

\subsection{Background}

\begin{newt}
As we leverage blockchain technologies and trusted execution environments to manage resource usage control, in the following we provide the necessary background information from these fields.
\end{newt}

\subsubsection{Data Access and Usage Control}
\label{sec:background:ucon}

Access control is a technique used to determine who or what can access resources in a computing environment~\citep{sandhu1994access}. In system infrastructures, access control is dependent upon and coexists alongside other security services. Such technologies require the presence of a trusted reference entity that mediates any attempted access to confidential resources. In order to decide who has rights to specific resources, access control frameworks make use of authorization rules, typically stored inside the system~\citep{koshutanski2003access}. A set of rules constitutes a policy. A popular approach of implementing access policies is through Access Control Lists (ACLs)~\citep{grunbacher2003posix}. Each protected resource has an associated ACL file, which lists the rights each subject in the system is allowed to use to access objects.

With the evolution of the web and decentralized data ecosystems, there is the need to move beyond managing access to resources via authorizations~\citep{akaichi2022usage}. Authorization predicates define limitations that consider the user and resource credentials and attributes. Usage control is an extension of access control whereby policies take into account obligations and conditions in addition to authorizations~\citep{lazouski2010usage}. Obligations are constraints that must be fulfilled by users before, during, or after resource usage. Conditions are environmental rules that need to be satisfied before or during usage.

One of the most highly cited usage control models is UCON$_{ABC}$~\citep{park2004uconabc}. The model represents policy rules by defining specific rights (e.g., operations to be executed) related to sets of subjects (e.g., users who want to perform an operation), objects (e.g., the resource to operate), authorizations, obligations, and conditions. \Newt{\emph{Attributes} are properties associated with subjects or objects.
UCON$_{ABC}$ improves conventional access control mainly through the following two concepts: (i) \emph{attribute mutability}, 
namely the change of attributes as a consequence of usage actions, and (ii) 
\emph{decision continuity}, i.e., the enforcing of policies not only as a check at access request time, but also during the subsequent resource usage.} 
Systems implementing usage control through the UCON$_{ABC}$ model require dedicated infrastructure to guarantee policy enforcement and monitoring in order to detect misconduct and execute compensation actions (e.g., penalties and/or right revocations).

The literature offers several alternative approaches that could potentially be used to represent usage control policies. For instance, \citet{hilty2007policy} propose a language named Obligation Specification Language (OSL) intended for distributed environments. \citet{bonatti2020machine} introduce the SPECIAL usage control policy language, which considers a policy as the intersection of basic entities governing data, processing, purposes, location, and storage of personal data. A comprehensive overview of existing usage control frameworks and their respective languages is provided by \citet{akaichi2022usage} and \citet{esteves2022analysis}.

\begin{newt}
The overarching goal of our work is to enable usage control in a decentralized environment. We provide a conceptual framework that serves as a blueprint for policy governance in a decentralized setting.
\end{newt}

\subsubsection{Trusted Execution Environments}
\label{sec:background:tee}
A Trusted Execution Environment (TEE) is a tamper-proof processing environment that runs on a separation kernel~\citep{mcgillion2015open}. Through the combination of both software and hardware features, it isolates the execution of code from the operating environment. The separation kernel technique ensures separate execution between two environments. TEEs were first introduced by~\citet{rushby1981design} and allow multiple systems requiring different levels of security to coexist on one platform. Thanks to kernel separation, the system is split into several partitions, guaranteeing strong isolation between them.
TEEs guarantee the authenticity of the code it executes, the integrity of the runtime states, and the confidentiality of the code and data stored in persistent memory. The content generated by the TEE is not static, and data are updated and stored in a secure manner. Thus, TEEs are hardened against both software and hardware attacks, preventing the use of even backdoor security vulnerabilities~\citep{DBLP:conf/trustcom/SabtAB15}.
There are many providers of TEE that differ in terms of the software system and, more specifically, the processor on which they are executed. In this work, we make use of the Intel Software Guard Extensions (Intel SGX)\footnote{\url{https://www.intel.co.uk/content/www/uk/en/architecture-and-technology/software-guard-extensions.html}. Accessed: \today.} TEE. Intel SGX is a set of CPU-level instructions that allow applications to create \emph{enclaves}. An enclave is a protected area of the application that guarantees the confidentiality and integrity of the data and code within it. These guarantees are also effective against malware with administrative privileges~\citep{zheng2021survey}. The use of one or more enclaves within an application makes it possible to reduce the potential attack surfaces of an application.
An enclave cannot be read or written to from outside. Only the enclave itself can change its secrets, independent of the Central Processing Unit (CPU) privileges used. Indeed, it is not possible to access the enclave by manipulating registers or the stack. Every call made to the enclave needs a new instruction that performs checks aimed at protecting the data that are only accessible through the enclave code.
The data within the enclave, in addition to being difficult to access, is encrypted. Gaining access to the Dynamic Random Access Memory (DRAM) modules would result in encrypted data being obtained~\citep{jauernig2020trusted}. The cryptographic key changes randomly each time the system is rebooted following a shutdown or hibernation~\citep{costan2016intel}.
An application using Intel SGX consists of a trusted and an untrusted component. We have seen that the trusted component is composed of one or more enclaves. The untrusted component is the remaining part of the application~\citep{zhao2016performance}. The trusted part of the application has no possibility of interacting with any other external components except the untrusted part. Nevertheless, the fewer interactions between the trusted and untrusted part, the greater the security guaranteed by the application.
\begin{newt}
Our work resorts to trusted execution environments to keep control of resources' utilization by enforcing the usage conditions set by data owners.
\end{newt}

\subsubsection{Decentralized Applications and Blockchain Oracles}
\label{sec:background:dapOracle}
With second-generation blockchains, the technology evolved from being primarily an e-cash distributed management system to a distributed programming platform for decentralized applications (DApps) ~\citep{mohanty2018ethereum}. Ethereum first enabled the deployment and execution of smart contracts (i.e., stateful software artifacts exposing variables and callable methods) in the blockchain environment through the Ethereum Virtual Machine (EVM)~\citep{buterin2014next}. 
The inability of smart contracts to access data that is not stored on-chain restricts the functionality of many application scenarios, including multi-party processes. Oracles solve this issue~\citep{blockchain:2016:wicsa}.

Oracles act as a bridge for communication between the on-chain and off-chain worlds. This means that DApps should also be able to trust an oracle in the same way it trusts the blockchain. Reliability for oracles is key~\citep{AFrameworkForBlockchain-BasedApplications,9086815}. 
Therefore, the designation and sharing of a well-defined protocol become fundamental for the proper functioning of the oracle's service, particularly when the oracles themselves are organized in the form of  networks for the interaction with decentralized environments~\citep{basile2021enhancing}.
As illustrated by~\cite{Muehlberger.etal/BPMBCF2020:FoundationalOraclePatterns}, oracle patterns can be described according to two dimensions: the information direction (inbound or outbound) and the initiator of the information exchange (pull- or push-based). While outbound oracles send data from the blockchain to the outside, inbound oracles inject data into the blockchain from the outside. Pull-based oracles have the initiator as the recipient, oppositely to push-based oracles, where the initiator is the transmitter of the information.
By combining the push-/pull-based and inbound/outbound categories, four oracle design patterns can be identified~\citep{oracle_pasdar}. 
A push-based inbound oracle (\emph{push-in} oracle for simplicity) is employed by an off-chain component that sends data from the real world. 
The push-based outbound (\emph{push-out}) oracle is used when an on-chain component starts the procedure and transmits data to off-chain components. 
The pull-based outbound (\emph{pull-out}) oracle is operated by an off-chain component that wants to retrieve data from the blockchain.
\Newt{Finally, the pull-based inbound (\emph{pull-in}) oracle enables on-chain components to retrieve information outside the blockchain}. 

\begin{newt}
We leverage the blockchain's tamper-proof infrastructure to record usage conditions associated with resources. We represent this information via smart contracts running in the blockchain and communicating with off-chain processes through oracles.
\end{newt}

\subsection{Related work}
\label{sec:rw}
Several works strive to provide more control and transparency with respect to personal data processing by leveraging blockchain distributed application platforms~\citep{DBLP:books/sp/XuWS19}. For instance, ~\citet{IoTBlock} defines an access control mechanism for IoT devices that stores a hash of the data in a blockchain infrastructure and maintains the raw information in a secure storage platform using a TEE. In the proposed framework, a blockchain based ledger is used in order to develop an audit trail of data access that provides more transparency with respect to data processing. \cite{10.1007/978-3-030-59013-0_30} propose a system, called PrivacyGuard, which gives data owners control over personal data access and usage in a data market scenario.

The literature offers numerous study cases in which usage control frameworks have been instantiated to increase the degree of privacy and confidentiality of shared data.
\citet{neisseUsage} propose a usage control framework in which a Policy Enforcement Point (PEP) keeps track of business operations and intercepts action requests while taking into consideration Policy Decision Point event subscriptions (PDP). \citet{bai2014context} addresses usage control in a Web Of Thing environment by adapting the UCON model for Smart Home ecosystems. \citet{8936349} introduce a secure usage control scheme for Internet of things (IoT) data that is built upon a blockchain-based trust management approach. While, ~\citet{DBLP:journals/winet/KhanZSAM20} conceptualizes a distributed usage control model, named DistU, for industrial blockchain frameworks with monitoring procedures that are able to revoke permissions automatically.

Additionally, there are several papers that propose frameworks or architectures that combine  blockchain platforms and decentralized web initiatives such as Solid web. \citet{10.1145/3366424.3385759} demonstrate how together Solid data stores (namely, \emph{pods}) and blockchains can be used for trustless verification with confidentiality. \citet{patel2019dauth} propose a fully decentralized protocol named DAuth that leverages asymmetric encryption in order to implement authentication. \citet{9064776} introduce a secure Solid authentication mechanism, integrating Rivest–Shamir–Adleman (RSA) signatures into permissioned blockchain systems. In turn, \citet{DBLP:conf/esws/BeckerVKBK21} demonstrate how data stored in Solid pods can be monetized by leveraging a blockchain based payment system. Whereas, ~\citet{havur2020greater} discuss how solid could potentially leverage existing consent, transparency and compliance checking approaches. 

Several studies have shown that blockchain and TEEs can profitably coexist. The state of the art proposes numerous cases where the combination of the two technologies leads to advantages in terms of data ownership, availability, and trust. One of these is the work of \citet{liang2017towards}, that propose a patient-centric personal health data management system with accountability and decentralization. The architecture of the framework employs TEEs to generate a fingerprint for each data access that are immutably maintained by a blockchain infrastructure. Whereas, \citet{lind2017teechain} designed and implemented a protocol named Teechain that integrates off-chain TEEs for secure and scalable payment procedures, built on top of the Bitcoin blockchain platform.


%
%
\section{Motivating scenario and requirements}
\label{sec:motivation}

The motivating use case scenario and the corresponding requirements, discussed in this section, are used not only to guide our work but also to contextualize theoretical notions introduced in the paper. 

\subsection{Motivating Scenario}
\label{sec:motivatingScenario}

\begin{figure}[t]
\centering
\includegraphics[width=9cm]{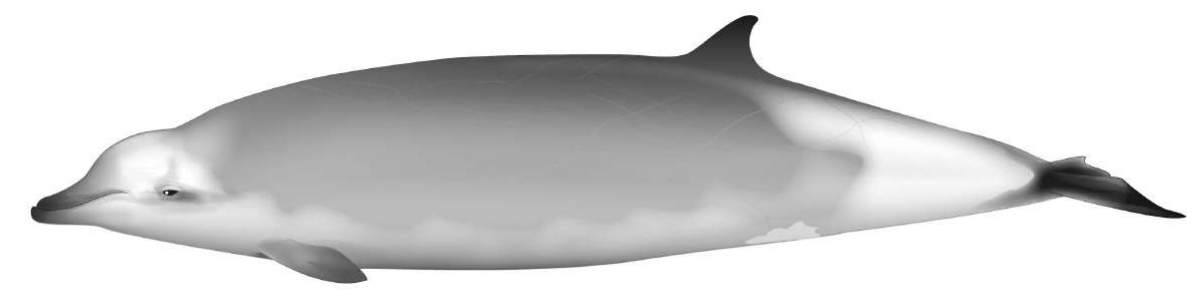}
\caption{A photographic representation of a Mesoplodon eueu~\citep{carroll2021speciation}. Image used under the Attribution 4.0 International (CC BY 4.0) license (\url{https://creativecommons.org/licenses/by/4.0/}). Cropped from original.}
\label{fig:mesoplodon}
\end{figure}

A new decentralized data market called DecentralTrading aims to facilitate data access across decentralized data stores. Alice and Bob sign up for the DecentralTrading market, pay the subscription fee, and set up their data nodes. Alice is a research biologist in the area of marine science and is conducting studies on deep ocean animals. Such species are difficult to identify due to the adverse conditions of their ecosystem and the lack of good-quality images. Bob is a professional diver with a passion for photography. He has collected several photos from his last immersion and the most scientifically relevant of them portrays a recently discovered whale species named `Mesoplodon eueu' showed in \cref{fig:mesoplodon}. 

Bob shares his photos with the DecentralTrading market by uploading them to his data node.
Once the images are shared, they can be retrieved by the other participants in the market. 
Moreover, he wants to establish rules regarding the usage of his images. \Cref{tab:example:constraints} illustrates the constraints he exerts on the data utilization, along with the \textbf{rule type} they represent (inspired by the work of \citealp{DBLP:conf/i-semantics/AkaichiK22}).
Bob makes his images available only for applications belonging to the scientific domain (this constraint belongs to the type of \textbf{domain rules}).
Moreover, he sets geographical restrictions by making the images usable only by devices located in European countries (\textbf{geographical rule}). Finally, Bob wants his photos to be deleted after a specific number of application accesses (\textbf{access counter rule}) or after a specific time interval (\textbf{temporal rule}). Therefore, he sets a maximum number of \num{100} local accesses and an expiry date of \num{20} days after the retrieval date. Bob gets remuneration from the DecentralTrading market, according to the number of requests for his resources. \Newt{At any point in time, Bob can ask the DecentralTrading market to get evidence that the rules associated with his image are being adhered to and check if there were attempts to use his image outside the specified rules}.

Bob's images of the Mesoplodon eueu species could be extremely useful for Alice's research, so she requests a specific picture of the gallery through her DecentralTrading node. Alice's node obtains a URL for Bob's node from the market and subsequently contacts Bob's node in order to retrieve a copy of the image, which is stored in a protected location of her device alongside the related usage rules. \Newt{Data shared in DecentralTrading is used by Alice and Bob through a set of known applications approved by the market community.} Alice opens the image through an app called `ZooResearch', which is used for the analysis of zoological images. \Newt{`ZooResearch' belongs to the set of approved applications, and it disables some tasks for data duplication by the operating system (OS) such as screenshots to replicate the image once it is accessed.} Since the domain of the application corresponds with the usage constraint set by Bob and her device is located in Ireland, the action is granted by Alice's node. \Newt{Afterwards, Alice tries to share the image through a social network application named `Socialgram', which also belongs to the set of supported applications.} Then, Alice's node denies the action since it goes against the application domain constraint set by Bob. Alice opens the image through `ZooResearch' \num{99} more times and, following the last attempt, the image is deleted from her node since the maximum number of local accesses of \num{100} has been reached. Therefore, Alice asks her DecentralTrading node to retrieve the image from Bob's node again. Since Alice starts working on a different research project, she stops using the Mesoplodon eueu's image. The image remains stored in the protected location of Alice's node until \num{20} days from the retrieval date have passed. Subsequently, Alice's node deletes the image from the protected location. 

\begin{table}[tb]
\caption{Schematization of the usage policy associated with Bob's `Mesoplodon.jpg' image. Every rule belongs to a rule type and consists of a subject, an action, an object, and a constraint.}
\label{tab:example:constraints}
\begin{center}
\footnotesize
\resizebox{\columnwidth}{!}{%
\tabulinesep=1mm
\begin{tabu}{|l|l|l|l|p{3cm}|} \hline
\backslashbox{\textbf{Rule type}}{\textbf{Rule components}}
& \textbf{Subject} & \textbf{Action}     & \textbf{Object} & \textbf{Constraint} \\ \hline
\textbf{Domain rule} & market members   & access the resource & Mesoplodon.jpg  & The resource can be processed only by research apps    \\ \hline
\textbf{Geographical rule} & market members   & access the resource & Mesoplodon.jpg  & The resource can be loaded only in European countries   \\ \hline
\textbf{Temporal rule} & market members   & access the resource & Mesoplodon.jpg  & The resource can be stored for up to \num{20} days              \\ \hline
\textbf{Access counter rule} & market members   & access the resource & Mesoplodon.jpg  & The resource can be opened up to \num{100} times       \\ \hline 
\end{tabu}
}
\end{center}
\end{table}

\subsection{Requirements}
The following concrete requirements are derived from our motivating scenario. The two top level requirements, which are inspired by the seminal work of \cite{akaichi2022usage}, are subdivided into more concrete sub-requirements.
\begin{itemize}
    \item[\textbf{(R1)}~] \textbf{Resource utilization and policy fulfillment must be managed by trusted entities.} According to \citet{akaichi2022usage}, a usage control framework must provide an enforcement mechanism that ensures usage policies are adhered to both before and after data are accessed. Therefore, the data market must be able to able to handle the access control and additionally the nodes of a decentralized environment must be equipped with a dedicated component managing the utilization of resources owned by other nodes.
    \begin{itemize}
        \item[\textbf{(R1.1)}~] \textbf{The trusted entity must be able to store resources obtained from other entities.} Once resources are accessed, they must be kept in a trusted memory zone directly controlled by the trusted entity. This requirement drastically reduces the risks of data theft or misuse. Considering our running example, it allows Alice to not only store Bob's resources but also to protect them from unauthorized access.
        \item[\textbf{(R1.2)}~] \textbf{The trusted entity must support the execution of programmable procedures that enforce constraints associated with resource usage.} Specific procedures must be designed in order to cater for the various usage policy rules types. The trusted entity must execute these procedures in order to enforce policies and control resource utilization. This aspect enables the logic associated with usage control rules, such as those defined in \cref{tab:example:constraints}, to be executed when Alice tries to use Bob's image.
        \item[\textbf{(R1.3)}~] \textbf{Resources and procedures managed by the trusted entity must be protected against malicious manipulations.} The trusted entity must guarantee the integrity of the resources it manages alongside the logic of the usage control procedures. Therefore, Alice should not be able to perform actions that directly manipulate Bob's image or corrupt the logic of the mechanisms that govern its use.
        \item[\textbf{(R1.4)}~] \textbf{The trusted entity must be able to prove its trusted nature to other entities in a decentralized environment.} Remote resource requests must be attributable to a trusted entity of the decentralized environment. Therefore, prior to Bob sending his image to Alice, it must be possible to verify that the data request has actually been generated by Alice's trusted node.
    \end{itemize}
    
    \item[\textbf{(R2)}~] \textbf{Policy compliance must be monitored via the entities of a governance ecosystem.} 
    According to \citet{akaichi2022usage}, usage control frameworks must incorporate a policy monitoring component. The monitoring, performed through one or more services, enables nodes to detect misconduct and unexpected or unpermitted usage. \Newt{This is, e.g., the mechanism thanks to which Bob can verify that Alice has never tried to open the picture of the Mesoplodon eueu with Socialgram.} 
    
    \begin{itemize}
        \item[\textbf{(R2.1)}~] \textbf{The governance ecosystem must provide transparency to all the nodes of the decentralized environment.}
        In order to gain the trust of the various nodes that comprise a decentralized environment, a governance ecosystem must guarantee transparency with respect to its data and procedures. This feature enables Bob to verify at any time that the usage policy associated with his image is being adhered to.
        \item[\textbf{(R2.2)}~] \textbf{Data and metadata maintained by the governance ecosystem must be tamper-resistant.}
        Once policies and resource metadata are sent to the governance ecosystem, their integrity must be ensured. The inability to tamper with resources and their metadata is crucial for the effective functioning of the governance ecosystem. Therefore, when Bob publishes images and their respective usage policies in the market, his node should be the only entity capable of modifying this metadata.
        \item[\textbf{(R2.3)}~] \textbf{The governance ecosystem and the entities that the form part of the ecosystem must be aligned with the decentralization principles.}
        It is essential that the governance ecosystem itself respects the decentralization principles, as centralized solutions would establish a central authority in which data and decisional power are accumulated. Hence, the monitoring functionality provided by the previously mentioned market scenario should not rely on centralized platforms and data stores. Bob's policies for the usage of the Mesoplodon eueu's photo are not uploaded on, nor verified by, any third-party service running on a specific server.
        \item[\textbf{(R2.4)}~] \textbf{The entities that form part of the governance ecosystem must be able to represent policies and verify their observance.}
        In order to provide monitoring functionality, entities in the governance ecosystem should be capable of managing usage policies. These entities should enact procedures for retrieving policy observance information directly from nodes that consume market resources.
        \Newt{This feature allows Bob to obtain evidence that Alice is using his image according to the rules stipulated in the usage policy and to detect any misbehavior}.
    \end{itemize}
    
    \end{itemize}

%
%
\section{Conceptual Resource Governance Framework}
\label{sec:framework}

\begin{figure}
\centering
\includegraphics[width=10cm]{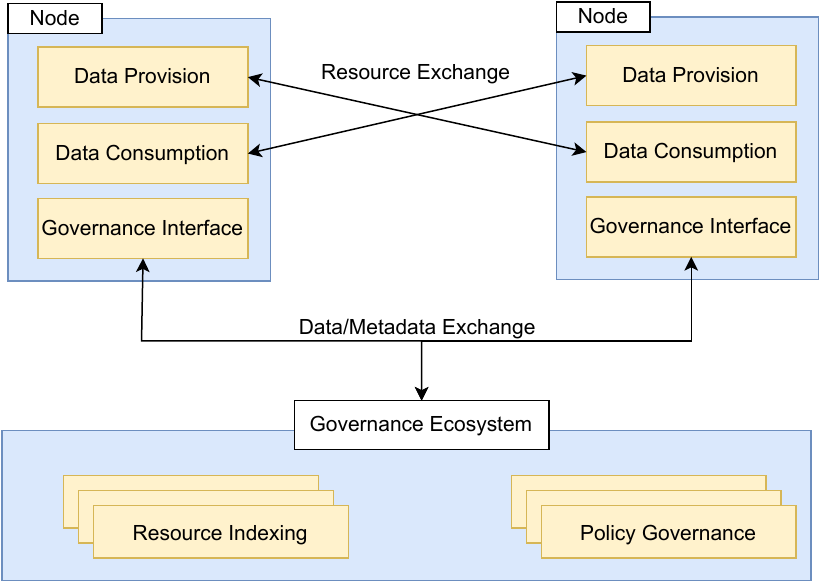}
\captionsetup{justification=centering}
\caption{High-level overview of the proposed conceptual resource governance (ReGov) framework.}
\label{fig:generalFramework}
\end{figure}

To cater for our motivating scenario and to meet the derived requirements, we propose a conceptual framework, named ReGov, that enables the governance of usage policies in decentralized web environments. ReGov generalizes the principles of data ownership and control, which constitute the foundations of numerous decentralized web initiatives. The ReGov framework extends these aspects by not only controlling data access but also supporting the continuous monitoring of compliance with usage policies and enforcing the fulfillment of usage policy obligations. The degree of abstraction of the ReGov framework means that it could potentially be instantiated in numerous decentralized web contexts. 

\subsection{ReGov Framework Entities}
According to the decentralization concept, the web is a peer-to-peer network with no central authority. In this scenario, data are no longer collected in application servers, but rather data are managed by nodes that are controlled by users (i.e., data owners determine who can access their data and in what context). Nodes communicate directly with other nodes in order to send and retrieve resources via the decentralized environment. 

\Cref{fig:generalFramework} depicts a high-level overview diagram of the ReGov framework. Nodes are characterized by the \Compo{data provision}, \Compo{data consumption}, and \Compo{governance interface} components. 
Governance ecosystems are responsible for indexing web resources, facilitating node and resource discovery, and monitoring resource usage. Thus, in our architecture, a \Compo{governance ecosystem} is constituted by the \Compo{resource indexing} and \Compo{policy governance} components.

\label{sec:architecture}

\begin{figure}[t]
\centering
\includegraphics[width=9cm]{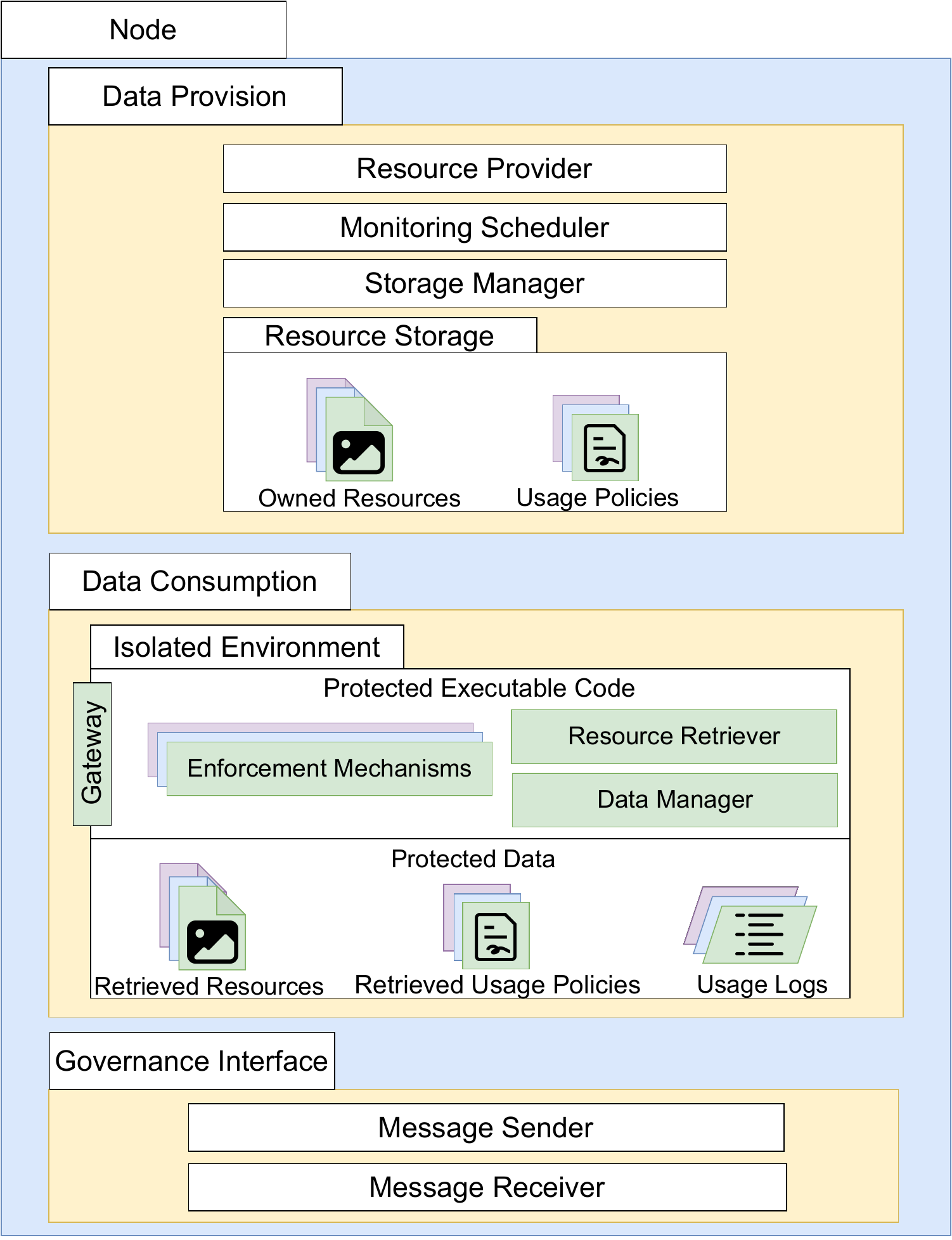}
\caption{Content of the data provision, data consumption and governance interface components.}
\label{fig:node}
\end{figure}

\subsubsection{Components of a Node}
\label{sec:framework:node}
A \Compo{node} is a combination of hardware and software technologies, running on user devices. As shown in \cref{fig:node}, each \Compo{node} comprises the following components.

\subsubsubsection{Data provision.}
The \Compo{data provision} component encapsulates the functionality that enable node owners to manage the sharing of their resources with other nodes in the decentralized environment. Users can interact with the \Compo{storage manager} to manually upload their data to the \Compo{resource storage} that is encapsulated within the \Compo{data provision} component. The upload operation also facilitates the definition of usage rules that are collected in usage policies associated with resources. Usage policies are represented in a machine-readable format (e.g., SPECIAL\footnote{\url{https://ai.wu.ac.at/policies/policylanguage/}. Accessed: \today.} and LUCON\footnote{\url{https://industrial-data-space.github.io/trusted-connector-documentation/docs/usage\_control/}. Accessed: \today.} policy languages) and stored in the \Compo{data provision} component alongside the resources. Additionally, when a new resource is uploaded, the \Compo{storage manager} forwards these rules and resource references to the \Compo{governance ecosystem}. In order to deliver the stored resources, the \Compo{data provision} component offers the logic for a \Compo{resource provider} that is capable of processing requests that allow other nodes to retrieve data. A data request must contain the necessary information to perform the authentication of the sender node. Therefore, the \Compo{resource provider} is able to authenticate resource requests to decide whether to grant or deny access to the requested resource based on the identity of the sender.  Several web service protocols could potentially be used to  implement the functionality offered by the \Compo{resource provider} (e.g., HTTP, FTP, Gopher). Once data are delivered, node owners can plan sessions to monitor the utilization of provisioned resources through the \Compo{monitoring scheduler}, which periodically forwards monitoring requests to the \Compo{governance ecosystem}.

Referring to our running example, Bob uses the functionality of the \Compo{storage manager} inside the \Compo{data provision} component to upload the images to his \Compo{node}. During the upload, he specifies the location where the images must be stored and the rules composing the images' \Compo{usage policy} (i.e. the image must be deleted 20 days after the retrieval date, the image can only be used in European countries). Therefore, these pieces of information are delivered to the \Compo{governance ecosystem}. The HTTP web service implementing the \Compo{resource provider} of Bob's \Compo{node} enables him to make his resource available to the other participants of the DecentralTrading market. The web service authenticates the requests for his images to determine whether the sender has the rights to access the resource. Finally, Bob can schedule monitoring sessions through the \Compo{monitoring scheduler}, in order to get evidence of the usage of his images by other nodes.

\subsubsubsection{Data consumption.} The \Compo{data consumption} component groups the functionalities that enable nodes to retrieve and use data in the network. \Compo{Data consumption} is built upon both hardware and software techniques that ensure the protection of sensitive data through an \Compo{isolated environment} that guarantees the integrity and confidentiality of protected data and executable code. The \Compo{isolated environment} contains the logic of a \Compo{resource retriever} that creates authenticable requests for data residing in other nodes. The \Compo{resource retriever} supports multiple web protocols (e.g., HTTP, FTP, Gopher) according to the implementation of the \Compo{resource provider} inside the \Compo{data provision} component. Therefore, if the \Compo{resource provider} is implemented as an FTP web service, the \Compo{request retriever} must be able to generate authenticable FTP requests. Once resources are retrieved alongside the related usage policies, they are controlled by the \Compo{data manager} that stores them in the \Compo{isolated environment}. To get access to a protected resource, local applications running in the \Compo{node} must interact with the \Compo{data manager} via the \Compo{gateway}, which acts as a bridge to the processes running in the \Compo{isolated environment}. The \Compo{gateway} is similarly employed when the \Compo{resource retriever} demands new resources from other nodes. In turn, \Compo{Enforcement mechanisms} governing data utilization are necessary to apply the rules of the usage policies. While controlling resources, the \Compo{data manager} cooperates with these mechanisms enabling the rules contained in the usage policies to be enforced. Each operation involving the protected resources is recorded in dedicated usage logs whose administration is entrusted by the \Compo{data manager} too. Usage logs facilitate policy monitoring procedures that employ these registers to detect potential misconduct.

As shown in the motivating scenario, Alice uses the \Compo{data consumption} component to get Bob's images, which she keeps in her own \Compo{node}. During the resource retrieval process, the \Compo{resource retriever} of Alice's \Compo{data consumption} component directly communicates with the \Compo{data provision} component of Bob's \Compo{node} through the \Compo{gateway}. After the retrieval, the image and the associated policy are maintained in the \Compo{isolated environment} and governed by the \Compo{data manager}. Considering the geographical rule, when Alice tries to open Bob's image with a local application, the app interacts with the \Compo{gateway}, which in turn, creates a communication channel with the \Compo{data manager}. The latter generates the execution of the \Compo{enforcement mechanism} of the geographical constraint. This mechanism consults the image's usage policy, retrieves the current geographical position of the \Compo{node}, and decides whether to grant the action. 

\subsubsubsection{Governance interface.} Nodes facilitate communication with the \Compo{governance ecosystem} via the \Compo{governance interface}. As we will see in~\cref{sec:architecture:interactions:monitoring}, messages flowing through the \Compo{governance interface} are crucial for resource usage monitoring. Indeed, the \Compo{governance ecosystem} can forward the interface messages such as requests for usage logs by remotely interacting with the \Compo{message receiver}. When a new message is received, the \Compo{governance interface} interacts with the other components of the \Compo{node} in order to deliver the information. Similarly, the \Compo{data provision} and \Compo{data consumption components} make use of the \Compo{message sender} to transmit data to the \Compo{governance ecosystem}. In order to provide continuous communication, the \Compo{governance interface} must constantly be active and listening for new messages.

\subsubsection{Components of the Governance Ecosystem}
\begin{figure}[t]
\centering
\includegraphics[width=\textwidth]{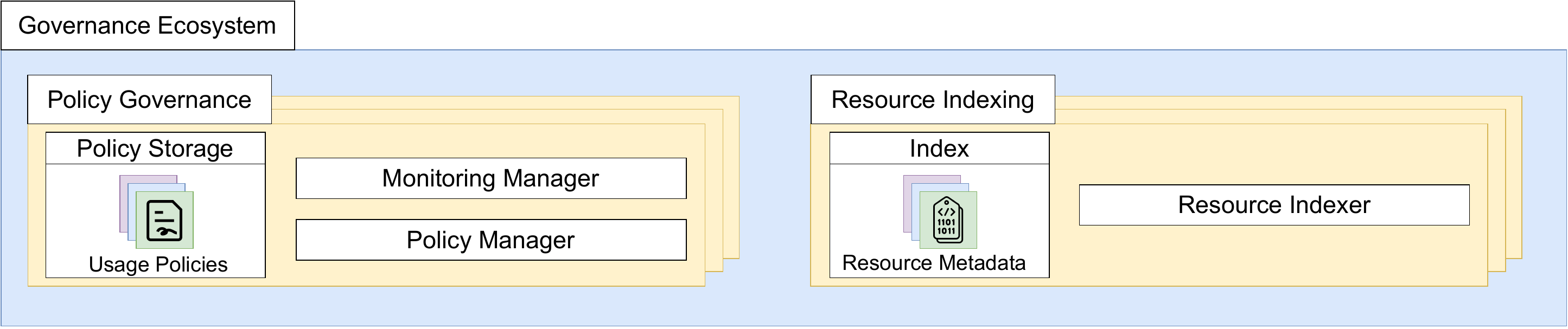}
\caption{Content of policy governance and resource indexing components inside the governance ecosystem}
\label{fig:governance_ecosystem}
\end{figure}

We extend the typical decentralized model by including the \Compo{governance ecosystem}, illustrated in \cref{fig:governance_ecosystem}. The ecosystem hosts the \Compo{resource indexing} and \Compo{policy governance} components, whose multiple instances are able to immutably store data and metadata, execute procedures, and communicate with all the nodes of the decentralized environment.

\subsubsubsection{Policy governance}. \Compo{Policy governance} components provide shared \Compo{policy storage} in which data owners publish applicable usage policies associated with resources. Policies are uploaded and modified through the \Compo{policy manager} of the component. In addition to their storage capabilities, \Compo{policy governance} components are able to execute procedures for policy monitoring. This function is supported by the \Compo{monitoring manager} of the component, containing the logic to verify the compliance of the policies stored inside the \Compo{policy storage}. Therefore, nodes forward monitoring requests to the \Compo{monitoring manager} which keeps track of resource usage and detects any illicit behavior. 

\subsubsubsection{Resource indexing}. Policies are associated with resources through \Compo{resource indexing} components. They contain metadata about the resources shared in the decentralized environment (e.g., identifiers, web references, owner node). When data owners upload new resources in their node, it interacts with the \Compo{resource indexer} of these components, in order to serialize the information of the shared data.

Referring to our running example, when Bob uploads his image to his \Compo{node} and specifies the corresponding usage rules in its policy, his \Compo{node} shares the image metadata (e.g., the HTTP reference \texttt{https://BobNode.com/images/Mesoplodon.jpg}) and the usage policy with respectively the \Compo{resource indexing} and \Compo{policy governance} components running in the \Compo{governance ecosystem}.
After Bob has delivered his `Mesoplodon.jpg' image to Alice's \Compo{node}, he can demand the verification of the image's utilization to the \Compo{policy governance} component holding the image's policy. The \Compo{policy governance} component retrieves the usage log of the image from Alice's device, by interacting with her \Compo{node}. Finally, Alice's usage can be verified based on the content of the usage log.

\subsection{Predominant ReGov Framework Operations}
Now that we have introduced the entities of our ReGov framework, we detail the predominant framework operations: data retrieval and monitoring. In the following, we simplify the processes by distinguishing owner nodes (i.e., nodes that are assuming the role of data providers) from data consumer nodes (i.e., nodes that are requesting access to and using resources), however, in practice, all nodes are dual purpose.

\begin{figure}[t]
\centering
\includegraphics[width=\textwidth]{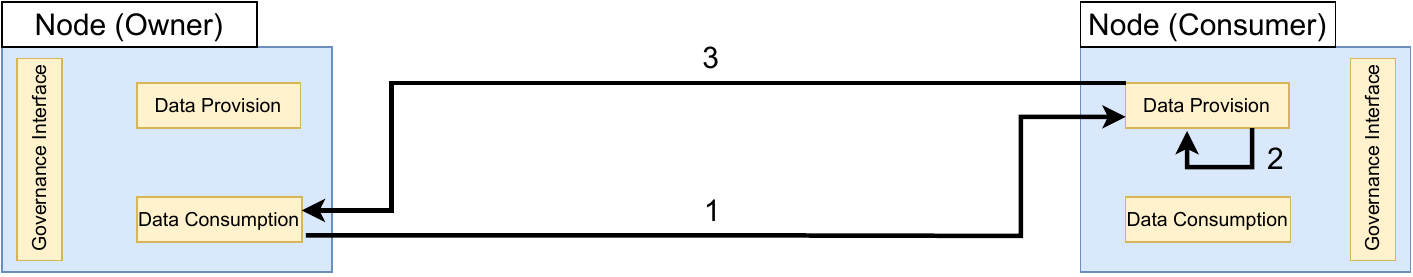}
\caption{Visualization of the ReGov framework data retrieval process.}
\label{fig:dataretrieval}
\end{figure}
\subsubsection{Data Retrieval}
The data retrieval process allows consumer nodes to retrieve a resource from the decentralized environment. \Cref{fig:dataretrieval} depicts a diagram representing the process. In order to obtain a specific resource, the data consumer \Compo{node} generates a new request and sends it to the owner \Compo{node}. We assume the consumer \Compo{node} already has the information needed to contact the owner node (e.g., IP address or web reference). This information can be obtained by reading resource metadata maintained by \Compo{resource indexing} components running in the governance ecosystem. The process starts when the \Compo{resource retriever} inside the \Compo{data consumption} component of the consumer \Compo{node} formats the request specifying the resource to be accessed and additional parameters intended for verification purposes. Subsequently, the request leaves the \Compo{isolated environment} through the \Compo{gateway} and is received by the \Compo{resource provider} inside the \Compo{data provision} component of the owner node (\textbf{1}). The latter uses the parameters of the request to verify the identity of the sender \Compo{node} (\textbf{2}). At this stage, the \Compo{resource provider} also verifies that the request has been generated in the \Compo{isolated environment} of a \Compo{data consumption} technology. Requests generated by alternative technologies are rejected. Once verified, the \Compo{resource provider} decides whether to grant access to the resource, according to the identity of the sender \Compo{node}. If access is granted, the resource provider interacts with the \Compo{storage manager} inside the \Compo{data provision} component in order to construct the response, which includes both the requested resource and its usage policy. Finally, the \Compo{resource retriever} of the consumer \Compo{node} obtains the resource, stores it in the \Compo{isolated environment} and registers it with the local \Compo{data manager} (\textbf{3}), as described in \cref{sec:framework:node}.

\subsubsection{Monitoring}
\label{sec:architecture:interactions:monitoring}
\begin{figure}[t]
\centering
\includegraphics[width=13cm]{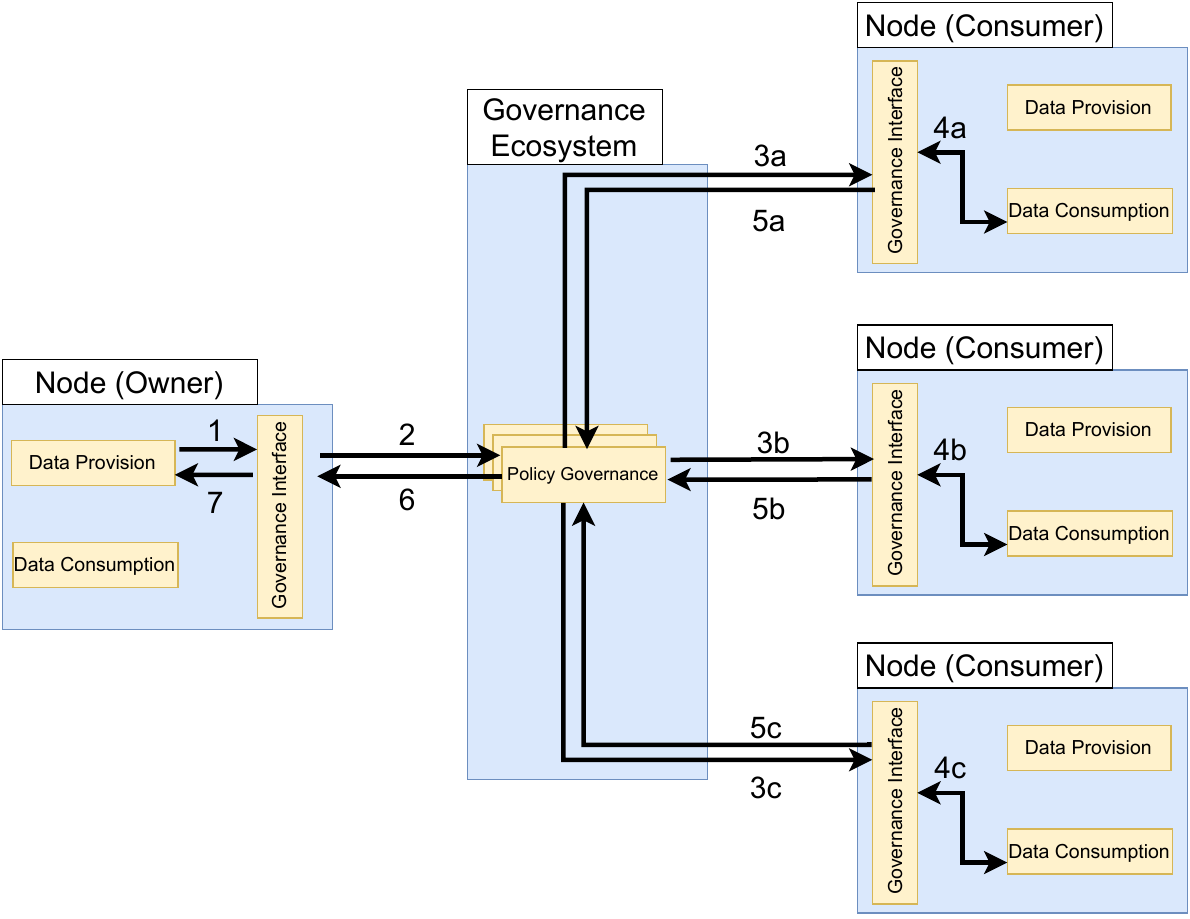}
\caption{Visualization of the ReGov framework data monitoring routine.}
\label{fig:monitoring}
\end{figure}

The policy monitoring process is used to continuously check if usage policies are being adhered to. In \cref{fig:monitoring}, we schematize the monitoring procedure. The owner node initiates the process via a scheduled job. Therefore the \Compo{monitoring scheduler} in the \Compo{data provision} component employs the \Compo{message sender} of the \Compo{governance interface} (\textbf{1}) to send a monitoring request, regarding a specific resource, to a \Compo{policy governance} component running in the \Compo{governance ecosystem} (\textbf{2}). Subsequently, the \Compo{policy governance} component forwards the request to provide evidence of utilization to each consumer \Compo{node} that has a copy of the resource (\textbf{3a, 3b, 3c}). In the depicted monitoring routine, we assume the resource whose usage must be monitored is held by three consumer nodes. In each of these nodes, the monitoring request is received by the \Compo{message receiver} of the \Compo{governance interface} that forwards, in turn, the request to the \Compo{data manager} running in the \Compo{isolated environment} inside the \Compo{data consumption} component (\textbf{4a, 4b, 4c}). The latter retrieves the usage log from the protected data storage and employs the \Compo{message sender} of the \Compo{governance interface} to forward the information to the \Compo{governance ecosystem}, which in turn ensures that all the consumer node responses are collected (\textbf{5a, 5b, 5c}). Finally, the evidence are returned to the \Compo{message receiver} (\textbf{6}) of the initiator \Compo{node}, which delivers the information to the \Compo{monitoring scheduler} (\textbf{7}).

%
%
\section{Blockchain and Trusted Execution Environment Instantiation}
\label{sec:BlockchainTEEinstantiation}
In this section, we describe an instantiation of the ReGov framework. To this end, we propose a prototype implementation of the DecentralTrading data market illustrated in the motivating scenario. The implementation integrates a trusted application running in a trusted execution environment and blockchain technologies to address usage control needs. The code is openly available at the following address: \url{https://github.com/ValerioGoretti/UsageControl-DecentralTrading}.

\begin{figure}[t]
\centering
\includegraphics[width=10cm]{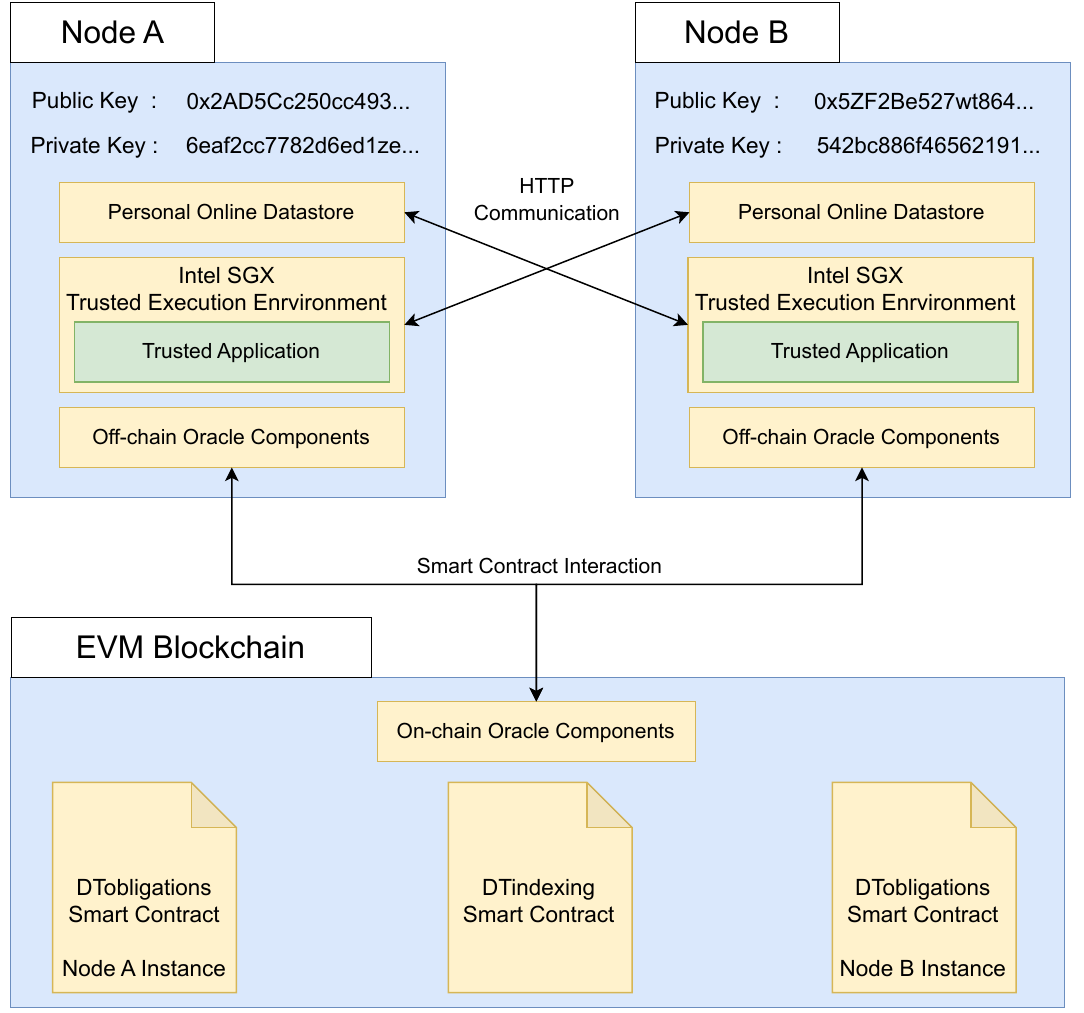}
\caption{High-level architectural overview of our ReGov framework instantiation.}
\label{fig:implementation}
\end{figure}

In \cref{fig:implementation}, we visualize the architecture of our ReGov framework instantiation. As shown in \cref{sec:framework}, the general framework assumes nodes of the decentralized environment are characterized by separate components dealing with \Compo{data provision} and \Compo{data consumption}. The \Compo{Data provision} functionality is implemented in a software component we refer to as a \Compo{personal online datastore}. We leverage security guarantees offered by the \Compo{Intel SGX trusted execution environment} in order to implement a \Compo{trusted application} containing the logic for \Compo{data consumption}. The \Compo{governance ecosystem} is realized by developing blockchain smart contracts that store information and execute distributed procedures. Our implementation involves an \Compo{EVM blockchain}\footnote{Ethereum Virtual Machine (EVM): \url{https://ethereum.org/en/developers/docs/evm/}. Accessed: \today.} (i.e., a blockchain based on the Ethereum Virtual Machine)  which hosts the \Compo{DTindexing} and \Compo{DTobligations} smart contracts. They fulfill the functions of the \Compo{resource indexing} and \Compo{policy governance} components of the general framework,  respectively. \Compo{DTindexing} is characterized by a unique instance managing the resource metadata of the decentralized environment. Instead, \Compo{DTobligations} is designed to be deployed multiple times. Therefore, each \Compo{node} is associated with a specific instance of this smart contract that stores the rules for its resources. The tasks performed by the \Compo{governance interface} are executed by blockchain oracles that provide a communication channel between the blockchain and the nodes of the decentralized environment. Oracles consist of \Compo{on-chain} components, running in the \Compo{EVM blockchain}, and \Compo{off-chain} components, operating within each \Compo{node}. We built the resource retrieval process between nodes using the HTTP communication standard. By interacting with smart contracts, nodes exchange metadata necessary for resource indexing and monitoring procedures.

\label{sec:instantiation}

Our implementation employs the asymmetric encryption methodology that underlies the \Compo{EVM Blockchain}, in order to provide an authentication mechanism for the environment's nodes. Each \Compo{node} is uniquely related to a public and private key pair that is used to sign authenticable data requests and transactions that transmit information to the blockchain and execute smart contract functions. \Newt{A private key is a 256-bit number generated through a secure random number generator. 
The corresponding public key is derived from the private key through the Elliptic Curve Digital Signature Algorithm \citep{johnson2001elliptic}.
The public key is connected to a unique account address on the \Compo{EVM Blockchain} derived as a 160-bit segment of the hash digest of the public key.
In our setting, \Compo{Node}s store their private key in an encrypted format to increase the degree of confidentiality of this information.
} 

In the following, we describe the technical details of the individual aspects of our implementation. In particular, we focus on features inherent to resource governance (data retrieval, enforcement, and monitoring) and avoid the implementation details related to the data market logic (e.g., subscription payments and remuneration mechanisms).

\subsection{Usage Policy Instantiation}
The first step of the instantiation process involves the definition of rule types that are used to stipulate usage policies. While our approach allows for a wide range of rules, we 
establish a specific subset of rules to demonstrate the capabilities of our ReGov framework.
In particular, we propose four types of rules inspired by the work of~\cite{DBLP:conf/i-semantics/AkaichiK22}. Each rule assumes that the target resource has already been retrieved and stored on the consumer device. In the following, we explain the various rule types that have already been introduced in the motivating scenario detailed in~\cref{sec:motivatingScenario}.

\subsubsubsection{Temporal rules.}
Through a temporal rule, data owners establish the maximum time a resource can be maintained within a consumer device. The rule is parameterized through an integer value representing the duration in seconds. Once the term expires, the rule stipulates that the resource must be deleted.

\subsubsubsection{Access counter rules.}
An access counter rule specifies a maximum number of local accesses that can be executed for a specific resource, after which, the resource must be deleted. The rule is parameterized with an integer value that defines the maximum number of accesses.

\subsubsubsection{Domain rules.}
The domain rule represents the purpose for which a resource can be opened. It is characterized by an integer value that identifies groups of applications that share the same domain. Known applications that are part of the domain group can execute local access to the resource.

\subsubsubsection{Geographical rules.}
A geographical constraint is a limitation on where a resource can be used. It is indicated by an integer code that specifies the territory in which the resource can be utilized.

\subsection{Personal Online Data Stores for Data Provision}
\begin{figure}[t]
\centering
\includegraphics[width=14cm]{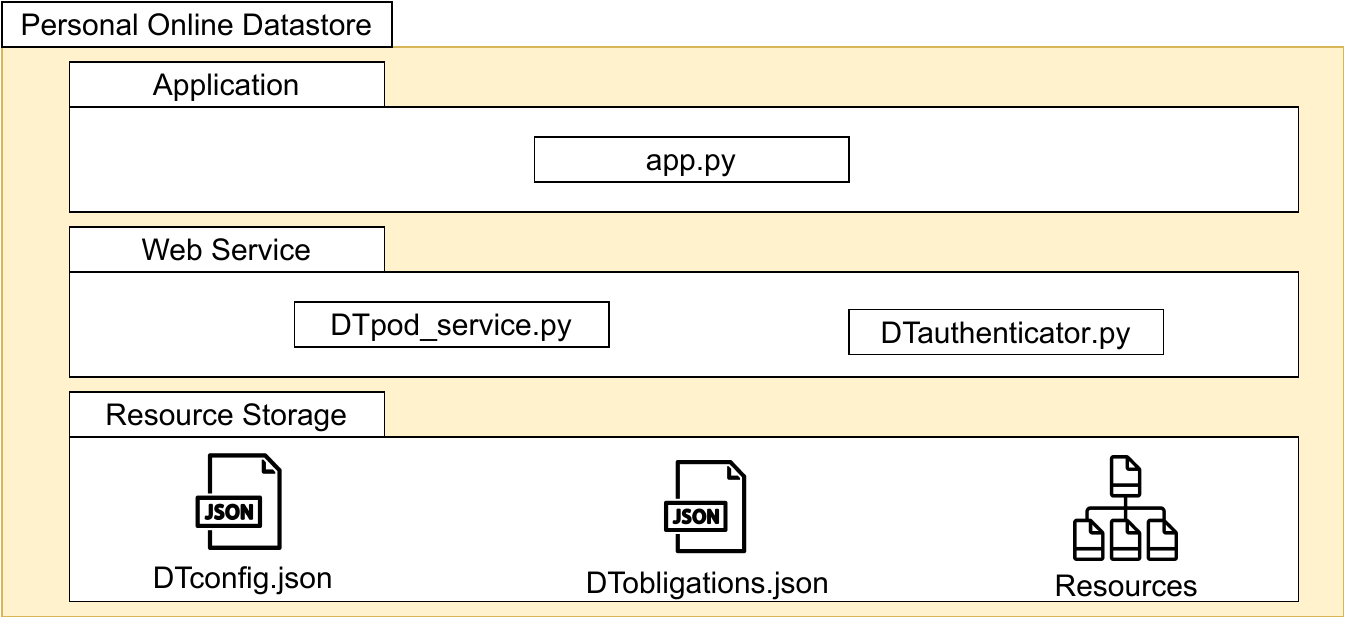}
\caption{Schematization of the personal online datastore implementation.}
\label{fig:personalOnlineDatastore}
\end{figure}

We develop the \Compo{personal online datastore} prototype using the Python language. Python's support for the Web3.py library\footnote{\url{https://web3js.readthedocs.io/en/v1.8.1/}. Accessed: \today.} enables the creation of communication protocols with the blockchain platform acting as the \Compo{governance ecosystem} of the decentralized environment. Our implementation also includes a graphical user interface developed with the Tkinter library\footnote{\url{https://docs.python.org/3/library/tk.html}. Accessed: \today.}.
As shown in~\cref{fig:personalOnlineDatastore}, our \Compo{personal online datastore} implementation is composed of three main parts: the \Compo{application}, the \Compo{web service} and the \Compo{resource storage}. The \texttt{app} module contains the executable code implementing the graphical user interface. 

\subsubsection{Resource Storage}
The resource storage contains the resources of the \Compo{personal online datastore}. The storage location is characterized by two meta-files named \texttt{DTconfig.json} and \texttt{DTobligations.json}. They contain descriptive and confidential information about the \Compo{personal online datastore} and its resources. \texttt{DTconfig.json} includes various attributes of a \Compo{personal online datastore}, such as its unique identifier, its node's public and private keys, the web reference to access data, and a list of the initialized resources. \texttt{DTobligations.json} holds rules that apply to the resources of the storage. The user can establish a default policy inherited by all resources in the \Compo{personal online datastore}, except those with specific policies. Mentioning our running example, Bob interacts with the \Compo{personal online datastore} application to upload the `Mesoplodon.jpg' resource in the `/images' location inside the storage. During this process, Bob can establish the rules associated with the image. The initialization of the image generates the metadata to be held in the \texttt{DTconfig.json} and \texttt{DTobligations.json} metafiles.

\subsubsection{Web Service}
\label{subsect:WebServiceImplementation}
\begin{figure}[t]
\centering
\includegraphics[width=9cm]{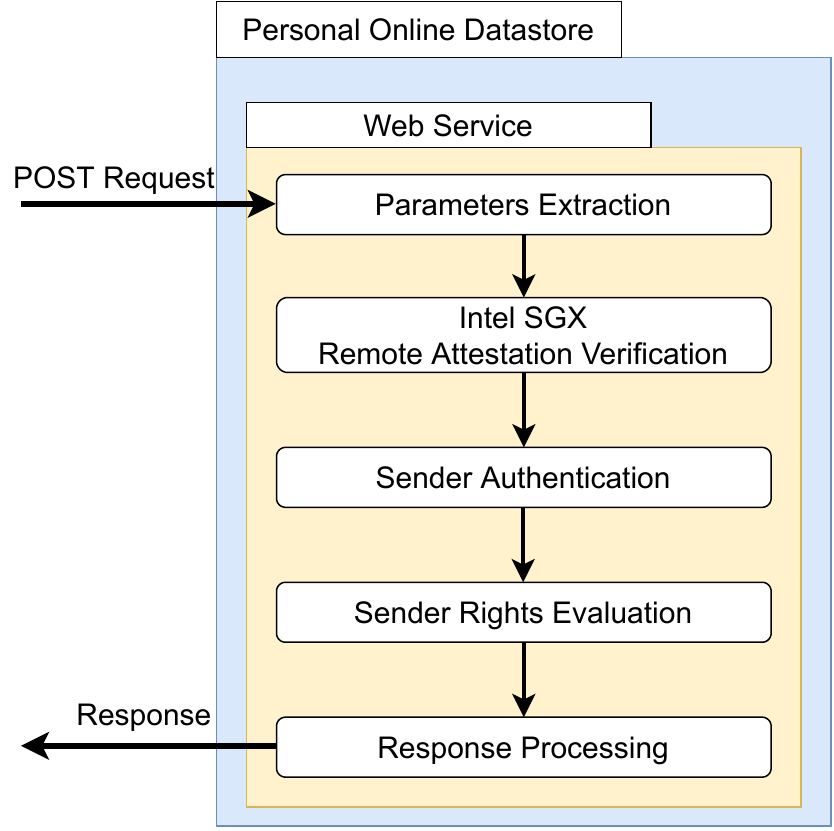}
\caption{Main stages of the ReGov data provision instantiation process.}
\label{fig:datastore:impl}
\end{figure}

The implementation of the data provision process is built upon the HTTP web standard. Our \Compo{Personal Online Datastore} prototype implements a \Compo{web service} that listens for HTTP requests, verifies the authenticity of the sender \Compo{node}, and delivers the requested data through HTTP responses. This approach enables the efficient and on-demand provision of initialized data. In~\cref{fig:datastore:impl}, we summarize the main stages of the data provision process, taking place in our \Compo{web service} implementation. 
The \texttt{DTpod\textunderscore service} Python class contains the core functionality for resource delivery. The class extends \texttt{BaseHTTPRequetsHandler} that enables the processing of GET and POST requests. Due to confidentiality reasons, the \Compo{web service} of the \Compo{personal online datastore} only responds to \Compo{POST request}s and ignores GET ones. The data provision process starts with the \Compo{Parameter extraction}, which takes place when a new \Compo{POST request} is received by the \Compo{web service}. The parameters inside the body of the \Compo{POST request} are crucial for the authentication and remote attestation procedures. In order to correctly demand a resource, requests must specify a URL composed of the web domain name of the service followed by the relative path of the requested resource inside storage. In the case of the motivating scenario, to retrieve Bob's image, Alice's node must generate an authenticable \Compo{POST request}, whose URL is `https://BobNode/images/Mesoplodon.jpg'.

Through remote attestation, the \Compo{web service} can verify that the resource request has been legitimately generated by a \Compo{Trusted Application} running a \Compo{Intel SGX trusted execution environment} of a \Compo{node}. Therefore, we leverage the \Compo{Intel SGX remote attestation verification} to establish a trusted communication channel between the consumer and the owner nodes. Once the attestation procedure ends successfully, the \Compo{web service} can be assured that the content of its response is managed by a \Compo{Data Consumption} technology inside the decentralized environment. 

\Compo{Sender authentication} takes place after the successful outcome of the remote attestation verification. The logic of our authentication mechanism is implemented through the \texttt{DTauthenticator} class, whose purpose is to use the \texttt{auth\textunderscore token} (a message signed with the sender's credentials) and \texttt{claim} (the public key of the sender) parameters inside the \Compo{POST request} to determine the sender \Compo{node}'s identity. Specifically, \texttt{auth\textunderscore token} refers to the URL of the resource to be accessed, encrypted with a private key. \texttt{DTauthenticator} is able to extract a public key from the \texttt{auth\textunderscore token} parameter when the request is received. If the extracted public key is equal to the \texttt{claim} parameter, the identity of the sender \Compo{node} is confirmed. At the end of the authentication procedure, Bob's \Compo{web service} identifies the sender of the request as Alice's \Compo{node}.

The determined identity is subsequently evaluated by the \Compo{web service} during the \Compo{Sender rights evaluation} to determine whether the consumer \Compo{node} can access the resource. Because our instantiation considers the decentralized environment related to the DecentralTrading data market (mentioned in \cref{sec:motivation}), this step establishes whether the sender \Compo{node} is associated with an active subscription (e.g., if Alice has an active subscription). However, the evaluation of alternative criteria, such as organization membership, can be freely integrated depending on the specific use case. In all cases, it is crucial to keep track of the consumer nodes that have accessed the \Compo{personal online datastore}'s resources by establishing their identity. 

Once the \Compo{POST request} has passed the necessary checks, the \Compo{Response processing} takes place. Therefore, the \Compo{web service} then interacts with the local storage to retrieve the requested resource, which, along with the associated policy, are inserted into the \Compo{response}.

\subsection{Trusted Execution Environment for Data Consumption}
\label{sec:implementation:tee}
\begin{figure}[t]
\centering
\includegraphics[width=15cm]{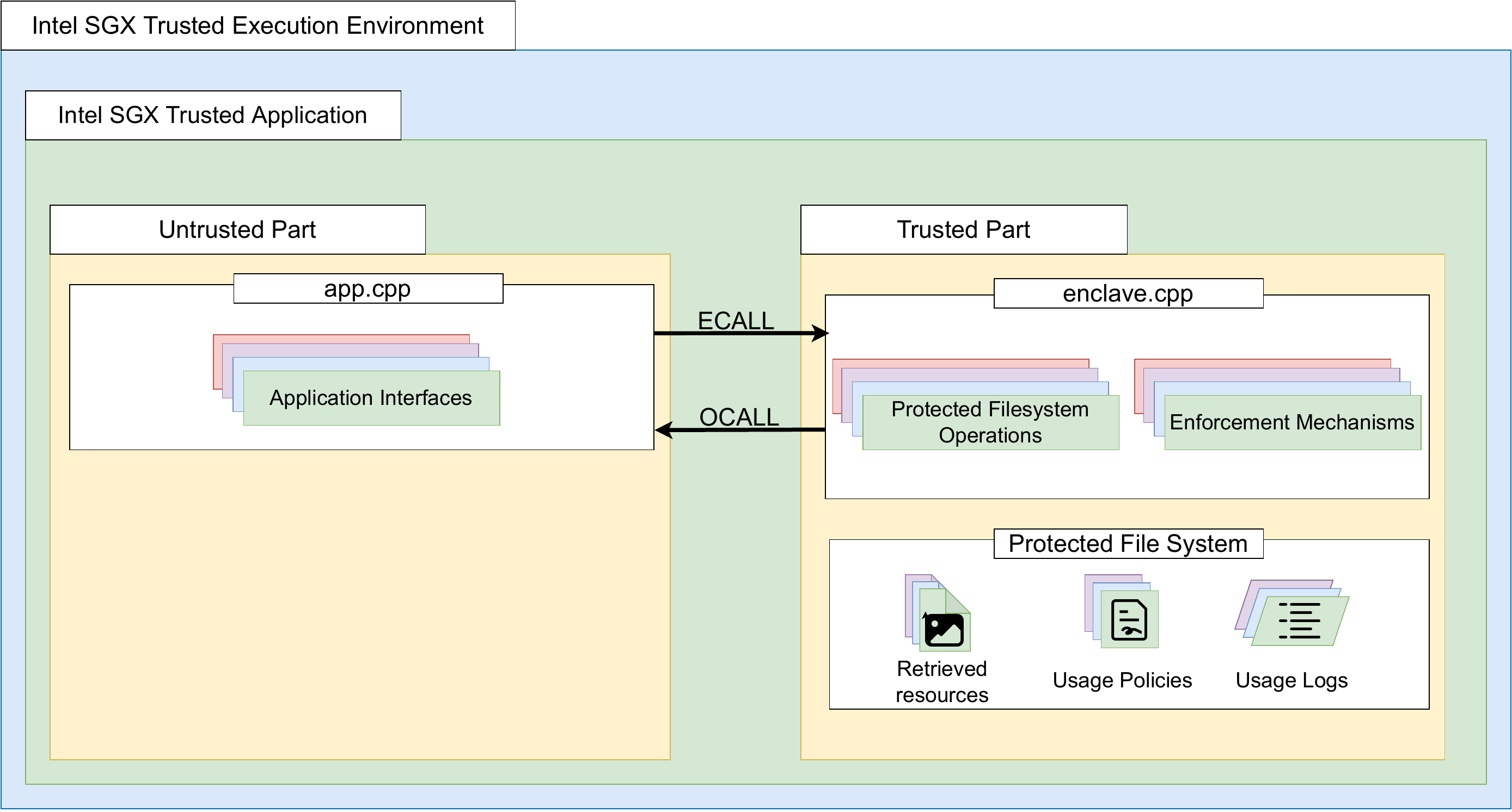}
\caption{Schematization of our trusted application composed of both trusted and untrusted elements.}
\label{fig:schemaTEE}
\end{figure}
\begin{newt}
The \Compo{trusted execution environment} manages the resources recovered within the consumer node. In~\cref{fig:schemaTEE}, we propose a schematization of our \Compo{trusted application} implementation. 
The trusted application consists of two fundamental components: the \Compo{trusted part} and the \Compo{untrusted part}. The \Compo{trusted part} comprises one or more enclaves. 
The \Compo{enclave}'s code is in the \texttt{enclave.cpp} file. It includes all the implementations of the \Compo{enforcement mechanisms} and a set of \Compo{protected file system operations} to handle the resources stored in it. The \Compo{trusted part} cannot communicate directly with the outside world. Any pieces of information that enter or leave the \Compo{trusted part} pass through the \Compo{untrusted part}. The \Compo{untrusted part}'s code is in the \texttt{app.cpp} file. This application has multiple \Compo{application interfaces} that are used to expose the application to the outside world. In order to communicate, the two parts use dedicated functions called \Compo{Ecall} and \Compo{Ocall}. `Ecall' stands for Enclave Call and represents an invocation made by a function in the \Compo{untrusted part} to the \Compo{enclave} (\Compo{trusted part}). The term `Ocall' (Out Call) refers to a call from the \Compo{enclave} to the \Compo{untrusted part}.
\end{newt}

\subsubsection{Data Protection}
The main purpose of using the \Compo{trusted application} is to manage and protect the data of other users obtained from the market. The \Compo{retrieved resources} are stored within the \Compo{enclave}, more specifically in its \Compo{protected file system}, because in this way they are decrypted only within the processor and only the enclave itself can access the processor in order to decrypt it. Within the enclave, both the \Compo{resources retrieved} by the user and the \Compo{usage policies} set by the owner are stored. Storing the \Compo{retrieved resource} within the \Compo{trusted part} is essential both from a data protection and a usage control perspective. In addition, the \Compo{usage policy} chosen by the data owner must also be saved in a secure space, as it could be tampered with by malicious code in order to be bypassed. 

\subsubsubsection{Protection of usage data}. When a user requests a piece of data, the request is received by the dedicated \Compo{application interface} in the \Compo{untrusted part}, and it is retrieved from the market. For instance, when Alice requests a photo of a Mesoplodon eueu from Bob, an identifier is assigned to this data before it is stored in the \Compo{enclave}. The identifier associated with the resource is used to index the retrieved resources and store them within the trusted part. A copy of the policies set by the owner, the rules set by Bob for the photo, is associated with it in order to store all the necessary resource information in the enclave. More specifically, when Alice wants to retrieve a piece of data from Bob, she interacts with the \Compo{untrusted part} and sends a post HTTP request to Bob's node. Within the request parameters, the resource in which the consumer is interested is specified, and an identifier is provided with which the consumer gets authenticated (as described in~\cref{subsect:WebServiceImplementation}). Finally, a certificate provided by Intel SGX Remote Attestation is added to the request, providing evidence that the request comes from a \Compo{trusted application}. Once the \Compo{personal online datastore} ensures that the other party involved in the communication is trusted, it sends the resource and policy information via an HTTP reply. Since the \Compo{trusted part} cannot communicate with the outside world, the response reaches the \Compo{untrusted part} who forwards it via an \Compo{ecall} to the \Compo{trusted part}. Once the resource arrives at the \Compo{trusted part}, it stores the data sent from the \Compo{personal online datastore} in the \Compo{enclave} using the \Compo{protected file system operations} that allow the \Compo{enclave} to manage the \Compo{protected file system}. Based on the example scenario, at this point the photo of the Mesoplodon eueu and the related \Compo{usage policies} set by Bob, the owner, are stored within Alice's \Compo{enclave}. 

\subsubsubsection{Protection of log data}. To keep track of the correct use of resources, all actions performed on them within the \Compo{trusted part} are stored in a \texttt{usage log file}. In short, all actions concerning the retrieved resources are stored. The objective is to let the data owner initiate a monitoring procedure through an oracle, 
to check whether resources are used in accordance with usage conditions.  When the \Compo{untrusted part} receives a monitoring request from the blockchain, it performs an \Compo{ecall} to request a copy of the \Compo{usage log} file stored in the \Compo{enclave} and returns it to the blockchain through an oracle to perform the monitoring. Referring to the example, all actions performed by Alice are recorded in a \Compo{usage log} file, and when Bob wants to check that everyone is using their resource correctly, he starts a monitoring procedure that aims to check all the \Compo{usage log} files of consumers who have retrieved the Mesoplodon eueu photos. When the \Compo{usage log} file is requested to be monitored, before sending a copy, the \Compo{trusted part} enters an entry to keep track of the monitoring request. 

\subsubsection{Implementation of the Enforcement Mechanisms}
\label{subsect:ChecksTEE}
In order to guarantee that data are accessed and used according to usage policies when a resource from the \Compo{trusted part} of a \Compo{trusted application} is requested by an external application, \texttt{enforcement mechanisms} must be implemented. These mechanisms are implemented within the \Compo{enclave} to ensure they are executed  within a \Compo{trusted environment}. 

\subsubsubsection{Receiving a request for access to a resource stored in the trusted application.} Before proceeding with the \Compo{enforcement mechanisms}, when the external application makes a request to the \Compo{trusted application}, the latter asks the external application to identify itself in order to check whether the sender is who it declares to be. More specifically, the \Compo{untrusted part} receives a request for access to a resource via the \Compo{application interfaces} and forwards it to the \Compo{trusted part} through an \texttt{Ecall} by invoking the \texttt{access\_protected\_resource} function, which verifies the identity of the claimant. Referring to the example, when Alice uses the `Zooresearch' or `Socialgram' applications, they have to authenticate themselves.

\subsubsubsection{Retrieval of the requested resource and its usage policy.} Once the external application has been authenticated, the \Compo{trusted application} gathers all the necessary information about it and accepts the request for the data that the external application is interested in and starts checking whether it is possible to access and use the resource. First, the \texttt{access\_protected\_resource} function retrieves the requested data and the associated policies, using the \texttt{get\_policy} function, set by the owner. Then, the \texttt{access\_protected\_resource} function invokes the different enforcement modules, passing the retrieved policies to it, in order to ensure that the rules are satisfied. In our implementation, four different enforcement modules have been developed. The proposed approach is highly flexible, thus catering for the extension of the existing rule types. The first mechanism in the enforcement process is checking the geographical position of the device.

\subsubsubsection{Geographical rule enforcement.} The \texttt{enforce\_geographical} function is invoked and passed the policy for the requested resource. The \texttt{get\_geo\_location} function (\texttt{Ocall}) is then used to retrieve the geographic location of the device from which the resource is being accessed. In the end, the geographic data set by the user and the current location are compared. If the position is correct, a positive result is returned to the \texttt{access\_protected\_resource} function, otherwise access is denied. Referring to the scenario, the \Compo{trusted application} uses Alice's location to check if it meets the location stipulated by Bob in his usage policy.

\subsubsubsection{Domain rule enforcement.} The \texttt{access\_protected\_resource} function invokes the \texttt{enforce\_domain} function by passing it the policy of the requested resource and information about the requesting application. Following a comparison between the application's domain and the domain set by the resource owner, if the domains are equal, the \texttt{enforce\_domain} function returns a positive result to the \texttt{access\_protected\_resource} function, which proceeds to the next check. Otherwise, access to the resource is denied. Looking at the example scenario, the domain of the application used by Alice is checked to determine if it satisfies the usage domain set by Bob. If Alice's application domain is correct, a positive result is returned. 

\subsubsubsection{Access counter rule enforcement.} The \texttt{enforce\_access\_counter} function is called by the \texttt{access\_protected\_resource} function with the policy for the requested resource. If the number of remaining accesses is greater than 1, the function decrements the maximum number of remaining accesses for that resource and returns with success to the \texttt{access\_protected\_resource} function. If the number of remaining accesses is equal to \num{1}, the function removes the resource and related policies from the \Compo{enclave} before returning a positive value, as the resource can no longer be accessed. In the motivating scenario, Bob set \num{100} as the maximum number of accesses to the resource. Each time Alice makes a request and logs in, the maximum number of hits left decreases. When the counter becomes \num{1}, Alice is allowed a last access to the Mesoplodon eueu's photo, and then the resource is deleted from her \Compo{trusted application}. Then, having successfully completed all the enforcement, the \texttt{access\_protected\_resource} function forwards the contents of the file to the \Compo{untrusted part}, which forwards it to the external requesting application. As already mentioned, all actions performed on the resources in the trusted application are saved on a \Compo{usage log} file, which keeps information and accesses made on the resources from when it is retrieved until it is deleted, maintaining an overview of the use of the resource. This \Compo{usage log} file makes it possible to prove and check that all resources have been used correctly within the trusted application.

\subsubsubsection{Temporal rule enforcement.} When it comes to temporal rules, the \Compo{untrusted part} periodically invokes the \texttt{Ecall} function called \texttt{enforce\_temporal} to verify that all resources within the trusted part have not expired. The \texttt{enforce\_temporal} function uses the \texttt{get\_trusted\_time} function to retrieve the current day. It then reads all resource policies stored within the \Compo{trusted part} and checks whether the date set on the policy is later than the current date. If a resource has expired, the \texttt{enforce\_temporal} function removes it. Each time this type of check is performed, it is written to the \Compo{usage log} file, and all deletions are also saved.

\subsection{Blockchain as a Governance Ecosystem}
In our instantiation, we leverage blockchain smart contracts in order to realize the \Compo{governance ecosystem}. Transparency, distribution, and immutability are the key features that make this technology highly suitable for our needs. The DecentralTrading implementation leverages the \Compo{EVM blockchain} platform hosting several interconnected smart contracts. \Compo{Node}s of the decentralized environment that are equipped with confidential blockchain public and private keys, sign authenticate transactions that generate the execution of smart contract functions. Processes that involve data exchange between \Compo{node}s and smart contracts are supported by blockchain oracles.

We implemented the smart contracts using the Solidity programming language\footnote{\url{https://docs.soliditylang.org/en/v0.8.17/}. Accessed: \today.}.
The smart contracts have been deployed in a local environment powered by the Ganache tool\footnote{\url{https://trufflesuite.com/ganache/}. Accessed: \today.} which enables the execution of a local blockchain replicating the Ethereum protocol and supporting the generation of transactions for testing purposes. In the following, we present the implementation details regarding the \Compo{DTindexing} and \Compo{DTobligations} smart contracts that fulfill the functionality of the \Compo{resource indexing} and \Compo{policy governance} components respectively. 

\subsubsection{DTindexing Smart Contract}
The \Compo{DTindexing} smart contract caters for the initialization of shared resources in the decentralized environment. The main goal of this component is to keep track of the decentralized environment’s data. Owner nodes interact with the smart contract to index their \Compo{personal online datastore}, sharing the necessary metadata for data retrieval. Consumer nodes make use of the smart contract to find references for registered resources through search functionality. \Cref{tab:classIndexing} represents the class diagram of the smart contract. The smart contract saves the following variables in the \texttt{Pod} struct in order to keep track of the information about personal online datastores:

\begin{lstlisting}[language=Solidity,numbers=none]
struct Pod { int id; address owner; bytes baseUrl; bool isActive; }
\end{lstlisting}
Similarly, the contract stores information about resources in a \texttt{Resource} struct, which consists of the following:

\begin{lstlisting}[language=Solidity,numbers=none]
 struct Resource{ int id; address owner; int podId; bytes url; bool isActive; }
\end{lstlisting}
The \texttt{Pod} and \texttt{Resource} structs are stored in the \texttt{podList} and \texttt{resourceList} array variables, respectively. The contract includes several methods for interacting with online datastores and resources, including the ability to register new ones, deactivate existing ones, and to search for them based on various criteria. For example, the \texttt{registerPod} method allows nodes to initialize new personal online datastores in the network. It takes as input a web reference for the online datastore service and the public key of the owner \Compo{node}. The function creates a new \texttt{Pod} struct and stores it in the \texttt{podList}. It also deploys a \Compo{DTobligations} smart contract (discussed next in detail), as every \Compo{personal online datastore} is related to one of these contracts. Finally, the function emits a \texttt{NewPod} event containing the identifier and the address of the \Compo{DTobligations} smart contract for the new online datastore. In our running example, Bob's node invokes this function to initialize his new \Compo{personal online datastore} providing the web reference \texttt{https://BobNode.com/} among the arguments. The function, in turn,  generates a new \texttt{Pod} struct. The \texttt{registerResource} method works similarly, generating a new \texttt{Resource} object and storing it in the \texttt{resourceList} state variable. In this case, Bob's \Compo{personal online datastore} employs this function to initialize the `Mesoplodon.jpg' image providing metadata such as the \texttt{https://BobNode.com/images/Mesoplodon.jpg} url. The \texttt{deactivateResource} and \texttt{deactivatePod} methods ensure that personal online datastores and resources are no longer accessible. Nodes submit metadata referring to new datastores and resources by using push-in oracles, that enable sending information to the blockchain. The smart contract also offers various search functions that can be useful for consumer nodes. The \texttt{getPodResources} method allows users to obtain a list of \texttt{Resource} structs stored in a specific datastore, identified by its integer identifier. The \texttt{getResource} method accepts an integer identifier as input and returns the \texttt{Resource} struct with that identifier. Referring to our use case scenario, Alice uses \texttt{getPodResources} to read the image's identifier that is given as a parameter to \texttt{getResource}, thanks to which the associated web reference is retrieved.

\begin{table}[]
\caption{Class diagram of the DTindexing smart contract.}
\label{tab:classIndexing}
\begin{center}
\centering
\scriptsize
\resizebox{0.8\columnwidth}{!}{%
\tabulinesep=0.7 mm
\begin{tabu}{|l|}
\hline
\multicolumn{1}{|c|}{\texttt{DTindexing}}                                                \\ \hline
\texttt{private podsCounter: int}                                                                                                                        \\
\texttt{private resourceCounter: int}                                                                                                                    \\
\texttt{private dtSubscription: int}                                                                                                                     \\
\texttt{private podList: Pod{[}{]}}                                                                                                                   \\
\texttt{private resourceList: Resource{[}{]}}                                                                                                            \\ \hline
\texttt{private searchByType(tp: PodType): Pod{[}{]}}                                                                                                  \\
\texttt{{\textless}{\textless}event{\textgreater}{\textgreater} NewPod(idPod: int, obgliationAddress: address)}                                            \\
\texttt{{\textless}{\textless}event{\textgreater}{\textgreater} NewResource(idResource: int)          }                                                    \\
\texttt{{\textless}{\textless}modifier{\textgreater}{\textgreater} validPodId(id: uint, owner: address) }                                                  \\
\texttt{public getMedicalPods(idSubscription: uint): Pod{[}{]}}                                                                                          \\
\texttt{public getSocialPods(idSubscription: uint): Pod{[}{]}    }                                                                                       \\
\texttt{public getFinancialPods(idSubscription: uint): Pod{[}{]}  }                                                                                      \\
\texttt{public registerPod(newReferene: bytes, podType: PodType, podAddress: address): int   }                                                           \\
\texttt{public registerResource(podId: int, newReferene: bytes, idSubscription: uint): int {\textless}{\textless}validPodId{\textgreater}{\textgreater}} \\
\texttt{public getPodResources(podId: int, idSubscription: int): Resource{[}{]} }                                                                        \\
\texttt{public deactivateResource(idResource: int): Resource {\textless}{\textless}validResourceId{\textgreater}{\textgreater}    }                      \\ \hline
\end{tabu}
}
\end{center}
\end{table}
\subsubsection{DTobligations Smart Contract}

 We use the \Compo{DTobligations} smart contract to model usage policies inside the blockchain environment and execute their monitoring. The architecture of the implementation assumes the deployment of multiple instances of the smart contract, one for each \Compo{personal online datastore} in the network. Each \Compo{DTobligations} smart contract is associated with a specific \Compo{personal online datastore} that is the only entity allowed to establish and manage the rules associated with the stored resources. As we showed in our motivating scenario, the architecture of our implementation assumes the deployment of a dedicated \texttt{DTobligations} instance containing the rules for Bob's \Compo{personal online datastore}. In \cref{tab:classDTobligations}, we propose the class diagram of the \Compo{DTobligations} smart contract.

The \Compo{DTobligations} smart contract includes four structs, each of which, models a specific rule: \texttt{AccessCounterObligation}, which restricts the number of resource accesses on a client device; \texttt{CountryObligation}, which imposes restrictions on the countries in which a resource can be used; \texttt{DomainObligation}, which specifies the purposes for which resources can be used; and \texttt{TemporalObligation}, which imposes a maximum duration for resource storage. These are stored in an \texttt{ObligationRules} struct, which can apply to a specific resource or to the entire \Compo{personal online datastore}. The smart contract includes functions that allow nodes to set default rules for their \Compo{personal online datastore} and related resources. For instance, the \texttt{addDefaultAccessCounterObligation} and \texttt{addDefaultTemporalObligation} are used to set rules that are inherited by all the resources of the \Compo{personal online datastore}. Similarly, functions such as \texttt{addAccessCounterObligation} and \texttt{addTemporalObligation} establish rules that are applied to a specific resource of the datastore. Referring to our running example, Bob's \Compo{personal online datastore} invokes the \texttt{addTemporalObligation} giving as input the `Mesoplodon.jpg' identifier and the integer value that describes the time duration of 20 days. The \texttt{onlyOwner} modifier ensures that certain functions can only be invoked by using the blockchain credentials associated with the smart contract's owner. It is applied to the functions for rule modification, which can be invoked only by the owner \Compo{node}. In this way, Bob is sure that modification of the rules can only be executed by his \Compo{personal online datastore}. 
\begin{table}[]
\caption{Class diagram of the DTobligations smart contract.}
\label{tab:classDTobligations}
\begin{center}
\centering
\scriptsize
\resizebox{\columnwidth}{!}{%
\tabulinesep=0.7 mm
\begin{tabu}{|l|}
\hline
\multicolumn{1}{|c|}{\begin{tabular}[c]{@{}c@{}}\texttt{DTobligations} \\ \texttt{{\textless}{\textless}extends {\textgreater}\textgreater \space 
Ownable}\end{tabular}}                                    \\ \hline
\texttt{dtIndexing: DTindexing}                                                                                                                                                            \\
\texttt{defaultPodObligation: ObligationRules }                                                                                                                                            \\
\texttt{resourcesObligation: mapping(int={\textgreater}ObligationRules)   }                                                                                                                \\ \hline
\texttt{{\textless}{\textless}modifier{\textgreater}\textgreater hasSpecificRules(resourceId: int)        }                                                                                \\
\texttt{{\textless}{\textless}modifier{\textgreater}\textgreater isValidTemporal(deadline: uint)  }                                                                                        \\
\texttt{{\textless}{\textless}modifier{\textgreater}\textgreater isTheResourceCovered(idResource: int)     }                                                                               \\
\texttt{public constructor(dtInd: address, podAddress: address)}                                                                                                                           \\
\texttt{public getObligationRules(idResource: int): ObligationRules {\textless}{\textless}isTheResourceCovered{\textgreater}{\textgreater}       }                                         \\
\texttt{public getDefaultObligationRules(): ObligationRules }                                                                                                                              \\
\texttt{public addDefaultAccessCounterObligation(accessCounter: uint) }                                                                                                                    \\
\texttt{public addDefaultTemporalObligation(temporalObligation: uint) {\textless}{\textless}isValidTemporal, onlyOwner{\textgreater}{\textgreater}         }                               \\
\texttt{public addDefaultCountryObligation(country: uint) {\textless}{\textless}onlyOwner{\textgreater}{\textgreater}}                                                                     \\
\texttt{public addDefaultDomainObligation(domain: DomainType) {\textless}{\textless}onlyOwner{\textgreater}{\textgreater}}                                                                 \\
\texttt{public addAccessCounterObligation(idResource: int, accessCounter: uint): ObligationRules {\textless}{\textless}isTheResourceCovered, onlyOwner{\textgreater}{\textgreater}   }     \\
\texttt{public addDomainObligation(idResource: int, domain: DomainType): ObligationRules {\textless}{\textless}onlyOwner, isTheResourceCovered{\textgreater}{\textgreater}  }              \\
\texttt{public addCountryObligation(idResource: int, country: uint): ObligationRules {\textless}{\textless}onlyOwner, isTheResourceCovered{\textgreater}{\textgreater}  }                  \\
\texttt{public addTemporalObligation(idResource: int, deadline: uint): ObligationRules {\textless}{\textless}onlyOwner, isTheResourceCovered, isValidTemporal{\textgreater}{\textgreater}} \\
\texttt{public removeAccessCounterObligation(idResource: int) {\textless}{\textless}onlyOwner, isTheResourceCovered, hasSpecificRules{\textgreater}{\textgreater}      }                   \\
\texttt{public removeTemporalObligation(idResource: int) {\textless}{\textless}isTheResourceCovered, onlyOwner, hasSpecificRules{\textgreater}{\textgreater}  }                            \\
\texttt{public removeDomainObligation(idResource: int) {\textless}{\textless}isTheResourceCovered, onlyOwner, hasSpecificRules{\textgreater}{\textgreater}  }                              \\
\texttt{public removeCountryObligation(idResource: int) {\textless}{\textless}isTheResourceCovered, onlyOwner, hasSpecificRules{\textgreater}{\textgreater}  }                             \\
\texttt{public removeDefaultTemporalObligation() {\textless}{\textless}onlyOwner{\textgreater}{\textgreater}}                                                                              \\
\texttt{public removeDefaultAccessCounterObligation() {\textless}{\textless}onlyOwner{\textgreater}{\textgreater}}                                                                         \\
\texttt{public removeDefaultCountryObligation() {\textless}{\textless}onlyOwner{\textgreater}{\textgreater}}                                                                               \\
\texttt{public removeDefaultDomainObligation() {\textless}{\textless}onlyOwner{\textgreater}{\textgreater}}                                                                                \\
\texttt{public withSpecificRules(idResource: int): bool     }                                                                                                                              \\
\texttt{public monitorCompliance() {\textless}{\textless}onlyOwner{\textgreater}{\textgreater}}                                                                                            \\ \hline
\end{tabu}
}
\end{center}
\end{table}

The main goal of the monitoring procedure is to retrieve evidence from consumer nodes attesting to the utilization of resources, whose policies are represented by the \texttt{DTobligations} instance. The smart contract implements the \texttt{monitorCompliance} function, solely invocable by the contract owner, to initiate the monitoring procedure. When the function is used, it interacts with a pull-in oracle, that is able to retrieve external information outside the blockchain. Therefore, the \Compo{DTobligations} smart contract communicates with the on-chain component of the oracle (i.e. smart contract named \texttt{PullInOracle}) by invoking its \texttt{initializeMonitoring} function. The oracle generates a new \texttt{MonitoringSession} struct instance that contains information about the current state of the session and aggregates the external responses. The same function emits a \texttt{NewMonitoring} event. The emission of the event is caught by the off-chain components of the oracle, running in consumer nodes, that forward to the \Compo{SGX Intel trusted application} the command to provide the usage log of the resources involved. Once the usage log is retrieved, the information contained within it are sent to the on-chain component of the oracle through its \texttt{\_callback} method. The function aggregates the responses from consumer nodes and updates the involved \texttt{MonitoringSession} instance each time it is called. Once all the responses are collected, they are returned to the \Compo{DTobligations} smart contract at the end of the process.
In our running example, the procedure is started by Bob's \Compo{personal online datastore} using the \texttt{monitorCompliance} function. Subsequently, Alice's \Compo{SGX trusted application} is contacted by the pull-in oracle and it is asked to provide the usage log of the `Mesoplodon.jpg' resource. Alice's response contains information such as the number of local accesses to the image or the time from its retrieval. The evidence provided by Alice's \Compo{SGX trusted application} is collected, together with pieces of evidence provided by other nodes in the network, by the pull-in oracle. Finally, the oracle forwards the logs to Bob's instance of \Compo{DTindexing}.

%
%
\section{Evaluation}
\label{sec:evaluation}
We evaluate the implementation of the ReGov framework by taking two distinct approaches. In the first part of this section we revisit the specific requirements usage control requirements that were derived from out motivating scenario. While, in the second part, we examine the security, privacy, and affordability of our implementation.

\subsection{Requirement Verification}
In this section, we discuss how the previously established requirements are satisfied by our ReGov instantiation, following the methodology described in the study of~\cite{terry2005requirements}. Through the discussion of the requirements, we contextualize the use of the trusted execution environment and the blockchain respectively in our architecture. Both requirements are composed of several sub-requirements that express various environmental and technological functions.

\subsubsection{(R1) Resource utilization and policy fulfillment must be managed by trusted entities}
The first requirement (\textbf{R1}) stipulates that \textbf{resource utilization and policy fulfillment must be managed by trusted entities}. We employ a trusted execution environment in order to develop a trusted application executable inside our nodes. We implemented it using Intel SGX, as explained in \cref{sec:implementation:tee}.
Our design and implementation choice allows us to satisfy the following sub-requirements:

\subsubsubsection{(R1.1) The trusted entity must be able to store resources obtained from other entities.} In the proposed  ReGov framework instantiation, all resources retrieved from the data market by the untrusted part of a node are passed to the trusted part of a node in order to store them within the enclave. For storage, we use an Intel SGX function, called Protected File System Library, which allows the management of files containing the resources retrieved within the enclave. We chose to store the data in the enclave because  any information stored in it is encrypted and decrypted solely by the enclave.

\subsubsubsection{(R1.2) The trusted entity must support the execution of programmable procedures that enforce constraints associated with resource usage.} When a resource stored within the enclave is requested, before retrieving it, the enclave we have implemented executes all the application procedures provided by the resource policy, invoking the necessary enforcement functions. The proposed enclave only allows access to the resource if at the end of the execution of all enforcement procedures, all of them have given a positive result. Otherwise, the resource is not returned and access is denied. It is worth noting that the enforcement mechanism within the trusted application is implemented in a modular way. Although our current implementation is limited to four rule types, this feature allows developers to easily extend our implementation with additional rule types based on their specific needs. 

\subsubsubsection{(R1.3) Resources and procedures managed by the trusted entity must be protected against malicious manipulations.} In the proposed ReGov implementation, we store resources within the enclave, because it is secure and protected from unauthorized access. The trusted part cannot communicate directly with the outside world and thus avoids interacting with malicious software. In addition, all code included and executed in the trusted part is, in turn, trusted, as it is not possible to use third-party libraries. The data stored within the enclave are encrypted. Therefore, a direct attack on the memory by malicious software would not be able to read the data.

\subsubsubsection{(R1.4) The trusted entity must be able to prove its trusted nature to other entities in a decentralized environment.} 
When it comes to interaction between nodes, in order to prove a node's trustworthiness, we employ the Intel SGX remote attestation within our trusted application. This advanced feature allows a node to gain the trust of a remote node. The provided attestation ensures that the node is interacting with a trusted application using an updated Intel SGX enclave.

\subsubsection{(R2) Policy compliance must be monitored via the entities of a governance ecosystem}

The second requirement (\textbf{R2}) stipulates that \textbf{policy compliance must be monitored through entities running in a governance ecosystem}. In our ReGov framework, we propose the adoption of a governance ecosystem that we instantiate on top of blockchain technology. In the following, we show the suitability of blockchain for this role by addressing each sub-requirement.

\subsubsubsection{(R2.1) The governance ecosystem must provide transparency to all the nodes of the decentralized environment.}
 By allowing all nodes to view the complete transaction history of the blockchain technology, we are able to ensure that each participant of the decentralized environment has equal access to information and is able to independently verify the accuracy and integrity of governance data. Additionally, we implement the policy management tasks via smart contracts, the code for which is made publicly available within the blockchain infrastructure. This enables nodes in the decentralized environment to be aware of the governance processes that are being executed.

\subsubsubsection{(R2.2) Data and metadata maintained by the governance ecosystem must be tamper-resistant.} 
Our solution involves the storage of resource metadata and usage policies in data structures that are part of smart contracts. Through smart contracts functions, we implement functionality that can be used to upload and modify stored data. We leverage the asymmetric key encryption mechanism of the blockchain environment to verify that data modifications are performed by authorized users. Once data and metadata of ReGov are validated in a blockchain block, we rely on the cryptographic structure underlying the blockchain to guarantee the integrity of published smart contracts and the information contained therein.

\subsubsubsection{(R2.3) The governance ecosystem and the entities that the form part of the ecosystem must be aligned with the decentralization principles.}
We fulfill the decentralization principles by proposing a blockchain-based architecture that is inherently decentralized. In our implementation, we publish data and metadata through a network of validators rather than a central authority. This ensures that no single entity has control over shared data and smart contracts that are distributed in the blockchain ecosystem. Through decentralization, we secure the fairness and integrity of policy management and prevent any single authority of the decentralized environment from having too much control or disproportionate decision-making power.

\subsubsubsection{(R2.4) The entities that form part of the governance ecosystem must be able to represent policies and verify their observance.} The majority of smart contract technologies are characterized by Turing-complete programming languages. We use the expressive power of smart contracts to implement data structures that can be used to represent usage policies and automate their monitoring. We facilitate the communication between smart contracts and off-chain nodes by integrating oracle technologies that implement the protocols for data-exchange processes. 

\subsection{Architecture Discussion}
In this section, we broaden our discussion on the effectiveness of the proposed decentralized usage control architecture with a particular focus on privacy, security, 
and affordability. The criteria the discussion is based on have been inspired by the work of~\cite{discussionCriteria}.

\subsubsection{Security}
\label{sec:security}
Several works already show how the decentralized model makes it more difficult for attackers to compromise data, as they would need to gain access to multiple nodes rather than just one central server \citep{raman2019challenges,web3.0}. As per the vast majority of decentralized web initiatives, our implementation preserves the security of data residing in nodes through the \Compo{personal online datastore} component, which performs authentication and rights evaluation procedures to prevent unauthorized access to sensitive information or resources. 

Our solution introduces new components into the decentralized model whose security should be discussed. 
The metadata stored in smart contracts (usage policies and resource indexes) are protected from unauthorized updates through the consensus mechanism of the blockchain platform and its distributed nature, which makes this information immutable. Moreover, the state of distributed applications running in this environment can only be changed by transactions marked by a digital signature. This feature guarantees that usage policy modifications can only be executed by authorized entities.

The \Compo{Intel SGX trusted execution environment} provides a separate ecosystem for the execution of a \Compo{trusted application} that manages resource utilization. It has already shown its effectiveness in terms of preventing the injection of malicious code coming from the operating system of the client's machine~\citep{DBLP:conf/trustcom/SabtAB15}, which could jeopardize the integrity of the stored resources and the local representation of usage policies. Moreover, we also leverage the security guarantees offered by this technology to establish a protected environment in which the enforcement of the usage policies is ensured, inside the consumer's node.

The monitoring process, thanks to which nodes get evidence of the utilization of their resources, involves the interaction between the \Compo{EVM blockchain} and consumer nodes. The procedure involves the exchange of confidential information, the integrity of which must be secured. Interactions between the involved components are managed via blockchain oracles that are capable of ensuring the legitimacy operations~\citep{trustworthyOracles}. By definition, oracles establish secure communication protocols that enable on-chain and off-chain computations to send and receive data safely.

\begin{newt}
Security and verification of data consumption are enforced by the ensemble of smart contracts, trusted execution environments, and remote attestations. Through the latter, data providers are able to remotely verify the integrity of a node's data consumption component and thwart attempts to instantiate malicious consumer nodes in the decentralized environment. Nevertheless, data provision of inappropriate information through published data is a practice that requires automated ex-post checking and whistleblowing~\citep{Kirrane.DiCiccio/AIChain2020:BlockConfess}.
\end{newt}

\begin{newt}
We remark that ReGov  cannot supervise users' actions outside the digital context of the decentralized environment. For example, it is unable to prevent users from taking a picture of a protected image resource using a separate camera, or copying reserved information displayed on the screen. The framework is intended to operate at the digital level. Therefore, ReGov monitors and controls data access, processing, and distribution, ensuring that it is utilized in compliance with the associated policy. Our motivating scenario resorts to a list of approved applications that guarantee fair data elaboration and facilitate misconduct uncovering. Considering the running example, applications like ``Socialgram'' put in place procedures that counteract OS screen recording actions. In addition, unfair activities that break the enforcement mechanism can be detected by the presented monitoring routines, enabling data owners to indict malicious users.
\end{newt}

\subsubsection{Privacy}
\label{sec:privacy}
Privacy is key for decentralized web environments trying to take personal data out of the control of single organizations. With usage control, users can benefit from a greater level of privacy, as they have a way to determine how their resources are being used. However, enforcement and monitoring mechanisms that characterize usage control involve the exchange of data and metadata whose confidentiality should constantly be guaranteed.

One of the most critical issues of our solution regarding confidentiality relates to the blockchain metadata, which are publicly exposed in smart contracts. Public blockchains, such as Ethereum, provide public ledgers, thus allowing every node of the decentralized environment to get access to usage policy and resource locations. \Newt{Despite the possibility of specifying private variables in smart contracts, the method invocations thanks to which those variables are set are recorded in publicly readable transactions. Therefore, blockchain users can freely deduce the state of a private variable by inspecting the public transactions associated with the invocation of the setter methods.} In some use cases, it may be desirable to keep this data public. However, there may also be a need to encrypt data stored in the blockchain, so that only authorized parties (those that have access to the decryption key) can read this metadata~\citep{pan2011survey,marangone2022fine}.

The confidentiality of the shared resources must be regulated after their retrieval inside consumer nodes, in order to apply the constraints associated with their policy rules. Our implementation leverage the \Compo{Intel SGX trusted execution environment} that manages retrieved resources through the \Compo{SGX Protected File System (PFS)}. One of the key features of SGX-PFS is that it allows for files to be stored in a secure, encrypted format, even when the operating system is not running. This makes it difficult for attackers to access the resources, as they would need to have physical access to the machine and be able to bypass the SGX hardware security features in order to read the contents of the files.

\subsubsection{Affordability}
\label{sec:affordability}
\begin{table}[]
\caption{Gas expenditure of the DTobligations and DTindexing smart contracts. Costs are expressed in Gas units.}
\begin{center}
\centering
\scriptsize
\resizebox{15 cm}{!}{%
\tabulinesep=1mm
\begin{tabu}{lS[table-format=7.0] l lS[table-format=7.0]}
\cline{1-2} \cline{4-5} 
\multicolumn{2}{c}{\textbf{DTobligations}}                                                  & \textbf{} & \multicolumn{2}{c}{\textbf{DTindexing}}                                  \\ \cline{1-2} \cline{4-5} 
{\textbf{Function}}                      & {\textbf{Cost}} & \textbf{} & {\textbf{Function}}    & {\textbf{Cost}} \\ \cline{1-2} \cline{4-5}  
{deployment}                             & 2057988       &           & {deployment}           & 3255000       \\
{\texttt{addDefaultAccessCounterObligation($\cdots$)}}    & 62627         &           & {\texttt{registerPod($\cdots$)}}        & 2082494       \\
{\texttt{addDefaultTemporalObligation($\cdots$)}}         & 62638         &           & {\texttt{registerResource($\cdots$)}}   & 143004        \\
{\texttt{addDefaultDomainObligation($\cdots$)}}           & 44219         &           & {\texttt{deactivateResource($\cdots$)}} & 21465         \\
{\texttt{addDefaultCountryObligation($\cdots$)}}          & 62561         &           & {}                     & {}              \\
{\texttt{addAccessCounterObligation($\cdots$)}}           & 138768        &           & {}                     & {}              \\
{\texttt{addTemporalObligation($\cdots$)}}                & 97737         &           & {}                     & {}              \\
{\texttt{addCountryObligation($\cdots$)}}                 & 97728         &           & {}                     & {}              \\
{\texttt{addDomainObligation($\cdots$)}}                  & 79452         &           & {}                     & {}              \\
{\texttt{removeDefaultAccessCounterObligation($\cdots$)}} & 23780         &           & {}                     & {}              \\
{\texttt{removeDefaultTemporalObligation($\cdots$)}}      & 16079         &           & {}                     & {}              \\
{\texttt{removeDefaultDomainObligation($\cdots$)}}        & 24747         &           & {}                     & {}              \\
{\texttt{removeDefaultCountryObligation($\cdots$)}}       & 23758         &           & {}                     & {}              \\
{\texttt{removeAccessCounterObligation($\cdots$)}}        & 28184         &           & {}                     & {}              \\
{\texttt{removeTemporalObligation($\cdots$)}}             & 28151         &           & {}                     & {}              \\
{\texttt{removeCountryObligation($\cdots$)}}              & 28173         &           & {}                     & {}              \\
{\texttt{removeDomainObligation($\cdots$)}}               & 38111         &           & {}                     & {}              \\
{\texttt{monitorCompliance($\cdots$)}}                    & 42000         &           & {}                     & {}              \\ \cline{1-2} \cline{4-5} 
\end{tabu}
}

\label{tab:gas}
\end{center}
\end{table}

The affordability of our solution is strongly related to the costs associated with the smart contracts running in the blockchain ecosystem. \Compo{EVM blockchain}s associate the execution of smart contracts with a fee charged to the invoking user, according to the complexity of the code to be executed. This fee is measured in (units of) Gas. In~\cref{tab:gas}, we collect the Gas expenses associated with the functions of the \texttt{DTobligations} and \texttt{DTindexing} smart contracts. The table omits their read functions, for which no transactions need to be sent to the network.

The deployment cost of \Compo{DTindexing} is \num{3255000} Gas units. The \texttt{registerPod} method is the most expensive \texttt{DTindexing}'s function (\num{2082494} Gas units) as it involves the deployment of a new contract instance, too. The Gas consumption of \texttt{registerResource} turns out to be significantly lower, requiring \num{143004} Gas units. The least expensive function of the smart contract is \texttt{deactivateResource} with an expenditure of \num{21465} Gas units. 

\Compo{DTobligations} is deployed during the registration of a new personal online datastore at the cost of \num{2057988} Gas units. \Compo{DTobligations} offers methods and functions to modify the obligation rules related to the resources contained in personal online datastore. Among the functions for adding rules, the most expensive one is \texttt{addAccessCounterObligation} with a value of \num{138768} Gas units. However, the adding of a domain restriction through \texttt{addDefaultDomainObligation} costs significantly less with \num{44219} Gas units per invocation. Methods for rule deactivation determine a lower expense than the previous ones. The cheapest among them is \texttt{removeDomainObligation} (\num{16079} Gas units). The cost required to initialize a monitoring process through the \texttt{monitorCompliance} function is \num{42000} units of Gas.

\Newt{As expected, operations involving new smart contract deployments are the most expensive ones. However, these costs are associated with one-time operations performed at setup time (at the bootstrapping of the platform, or every time a new pod is registered). On the other hand, functions intended for more frequent invocations (e.g., to monitor compliance or update rules)
are characterized by significantly lower costs. Costs in fiat money are subject to high variability, as they depend on multiple factors including the network capacity utilization, the price in cryptocurrency per Gas unit, and the market exchange rate of the cryptocurrency. Also, these values change depending on the EVM blockchain in use (e.g., Ethereum\footnote{\url{https://ethereum.org/}. Accessed: \today.},
Avalanche\footnote{\url{https://www.avax.com/}. Accessed: \today.}, Polygon\footnote{\url{https://polygon.technology/}. Accessed: \today.}, and more).
At the time of writing, we empirically found variations of four orders of magnitude%
\footnote{%
\begin{newt}%
The amount of gas needed for the deployment of the \texttt{DTindexing} smart contract, e.g., is \num{3255000}. During our experiments, the price per Gas unit in the Ethereum public network amounted to \num{36.15} Gwei (one GWei is worth $10^{-9}$~ETH). The ETH/EUR exchange rate was $1/1590$ EUR. The total gas cost price was thus \num{187.09} EUR.
Other EVM blockchains exhibited lower Gas prices or exchange rates, decreasing the overall cost in fiat money. Considering the Avalanche and Polygon platforms, their Gas price was \num{42.56}~and~\num{168.65} Gwei, respectively. The AVAX/EUR exchange rate was $1/15.67$, and the MATIC/EUR exchange rate was $1/1.19$. As a result, the total expenses amounted to \num{2.17}~and~\num{0.65}~EUR, respectively.
Data collected: 14 March 2023, 11:30 pm. Our smart contract deployments can be found on the G\"orli Ethereum test network at \url{https://goerli.etherscan.io/address/0xb0fe7d07947d9dd7cda47825e61ec14b98ef271a}, on the Fuji Avalanche test network at \url{https://testnet.snowtrace.io/address/0x0082698263ccc5765c97404af39023daefe20096}, and on the Mumbai Polygon test network at \url{https://mumbai.polygonscan.com/address/0x9ee2cb5ef7b1449d615d9fd0f9b167543e0d28eb}.%
\end{newt}%
}%
.
However, we remark that our implementation costs align with ERC721 implementations%
\footnote{\url{https://eips.ethereum.org/EIPS/eip-721}. Accessed: \today.}%
.
For example, the deployment fees of the Ethereum Name Service (ENS)%
\footnote{\url{https://etherscan.io/token/0xc18360217d8f7ab5e7c516566761ea12ce7f9d72}. Accessed: \today.}, a non-fungible token in the neighboring area of personal information indexing, amounts to \num{2443978} Gas units%
\footnote{\url{https://etherscan.io/tx/0xff3ee18523c9ec20e62d31d3d3ce3e8bf25f5ffcdfc4c32cd43ed0a786cc8640}. Accessed: \today.%
}%
. 
The market scenario can support the structural expenses associated with the proposed implementation and provides an incentive system that allows users to earn money by sharing their data. However, cost reduction practices are necessary to increase usability. These include design improvements to the implementation's architecture as well as the adoption of side-chains and layer-2 networks.}

%
%
\section{Conclusion}
\label{sec:conclusion}

Since its inception, the web has evolved from a read-only medium for information dissemination to a ubiquitous information and communication platform that supports interaction and collaboration globally. Although the web is by design decentralized and thus is not controlled by any single entity or organization, the web as we know it today is dominated by a small number of centralized platforms. Consequently, the decentralized web initiative aims to promote research into tools and technologies that give data owners more control over their data and enable smaller players to gain access to data, thus enabling innovation. 

In this paper, we focus specifically on resource governance in a decentralized web setting. We extend the state of the art by proposing a conceptual resource governance framework, entitled ReGov, that facilitates usage control in a decentralized setting, with a particular focus on policy respecting resource utilization and resource indexing and continuous monitoring. In order to demonstrate the potential of our ReGov framework, we propose a concrete instantiation that employs a trusted execution environment to cater for the former, and blockchain technologies to facilitate the latter. The effectiveness of the ReGov framework and our particular instantiation is assessed via a detailed analysis of concrete requirements derived from a data market motivating scenario and an assessment of the security, privacy, and affordability aspects of our proposal. 

Future work includes extending our primitive rule syntax to encompass more expressive usage control policies that are based on standard policy languages. Additionally, we plan to explore strategies for reducing the costs associated with the smart contracts running in the blockchain ecosystem. Studying incentivization mechanisms to encourage users to use the platform and possibly gain rewards for sharing information also paves the path for future endeavors.
\begin{newt}
The community-based categorization of applications interfaced with ReGov is a challenging aspect, the solution to which potentially involves the adoption of dedicated smart contracts for voting and arbitrage mechanisms. Also, erroneous or malicious misuse of ReGov such as the publication and disclosure of otherwise private information is beyond the reach of ReGov and would entail ex-post patrolling of the system. Studying these integrations with our framework is a task we envision for future work.
\end{newt}
Finally, we aim to conduct case studies with users to evaluate our approach in real-world settings.

\subsubsection*{Acknowledgments.}
The work of D.\ Basile, C.\ Di Ciccio, and V.\ Goretti was partially funded by the Italian Ministry of University and Research under grant ``Dipartimenti di eccellenza 2018-2022'' of the Department of Computer Science at Sapienza, by the EU-NGEU NRRP MUR under grant PE00000014 (SERICS), by the Cyber 4.0 project BRIE, and by the Sapienza project ``Drones as a Service for First Emergency Rresponse''. The work of S.\ Kirrane was funded by the FWF Austrian Science Fund and the Internet Foundation Austria under the FWF Elise Richter and netidee SCIENCE programmes as project number V 759-N.

\bibliographystyle{Frontiers-Harvard} 
\bibliography{bibliography}

\end{document}